\documentclass[12pt,letterpaper]{article}
\usepackage[latin1]{inputenc}
\usepackage{amsmath}
\usepackage{amsfonts}
\usepackage{amssymb}
\usepackage{graphicx,color}
\usepackage{color}
\usepackage{young}
\usepackage[vcentermath]{youngtab}
\usepackage{cite}

\usepackage{soul,xcolor}

\newcommand{\bet}{\beta}

\newcommand{\al}{\alpha}
\newcommand{\ga}{\gamma}
\newcommand{\de}{\delta}
\newcommand{\ep}{\epsilon}

\newcommand{\si}{\sigma}

\newcommand{\be}{\begin{equation}}
\newcommand{\ee}{\end{equation}}
\newcommand{\bea}{\begin{eqnarray}}
\newcommand{\eea}{\end{eqnarray}}

\begin{document}

\thispagestyle{empty}

\setcounter{page}{0}

\mbox{}
\vspace{-35mm}

\begin{center} {\bf \Large  Twisted Self-Duality for Linearized Gravity in $D$ Dimensions} 

\vspace{1.1cm}

Claudio Bunster$^{1,2}$,  Marc Henneaux$^{1,3}$ and Sergio H\"ortner$^3$

\footnotesize
\vspace{.4 cm}

${}^1${\em Centro de Estudios Cient\'{\i}ficos (CECs), Casilla 1469, Valdivia, Chile}

\vspace{.1cm}

${}^2${\em Universidad Andr\'es Bello, Av. Rep\'ublica 440, Santiago, Chile}

\vspace{.1cm}

${}^3${\em Universit\'e Libre de Bruxelles and International Solvay Institutes, ULB-Campus Plaine CP231, B-1050 Brussels, Belgium} \\

\vspace {9mm}

\end{center}
\centerline{\bf Abstract}
\vspace{.5cm}
The linearized Einstein equations in $D$ spacetime dimensions can be written as twisted self-duality equations expressing that the linearized curvature tensor of the graviton described by a rank-two symmetric tensor, is dual to the linearized curvature tensor of the ``dual graviton" described by a tensor of $(D-3,1)$ Young symmetry type.  In the case of $4$ dimensions, both the graviton and its dual are rank-two symmetric tensors (Young symmetry type $(1,1)$), while in the case of $11$ space-time dimensions relevant to $M$-theory, the dual graviton is described by a  tensor of $(8,1)$ Young symmetry type.

We provide in this paper an action principle that yields the twisted self-duality conditions as equations of motion, keeping the graviton and its dual on equal footing. 

 In order to construct a local, quadratic, variational principle for the twisted linear self-duality equations, it is necessary to introduce two ``prepotentials". These are also tensors of mixed Young symmetry types and  are obtained by solving the Hamiltonian constraints of the Hamiltonian formulation either of the Pauli-Fierz action for the graviton or of the Curtright action for its dual, the resulting actions being  identical. The prepotentials enjoy interesting gauge invariance symmetries, which are exhibited and generalize the gauge symmetries found in $D=4$.
 
 A variational principle where the basic variables are the original Pauli-Fierz field and its dual can also be given but contrary to the prepotential action, the corresponding action is non-local in space -- while remaining local in time. 
 
 We also analyze in detail the Hamiltonian structure of the theory and show that the graviton and its dual  are canonically conjugate in a sense made precise in the text.

\vspace{.8cm}
\noindent

\newpage


\section{Introduction}
\setcounter{equation}{0}

Electric-magnetic duality has emerged as a very interesting symmetry of a series of theories of increasing complexity \cite{Deser:1976iy,Deser,Henneaux:1988gg,ScSe,Deser:1997mz,Bunster:2011aw,Bunster:2011qp,Henneaux:2004jw,Deser:2004xt,Julia:2005ze,Hillmann:2009zf,Bunster:2012jp}. In all cases two key properties keep reappearing: (i) The symmetry is an invariance of the action, and not just of the equations of motion  \cite{Comment}) and (ii) The symmetry can be made manifest, while keeping the formulation simple, only at the price of giving up manifest spacetime covariance \cite{Others}. The latter property exhibits a  fascinating ``complementarity" between duality and spacetime covariance, which is further put in evidence by the fact that duality can be shown to imply Lorentz invariance, at least in the simple case of an abelian gauge field  \cite{Bunster:2012hm}. At a more technical level two general features that also appear are: (iii) The need to  reformulate the theory in terms of new variables (``prepotentials"), and (iv) A corresponding doubling of the gauge symmetry group. 

In the case of gravitational theories,  it is generally expected that  gravitational duality holds the key for exhibiting the conjectured infinite-dimensional Kac-Moody algebras (or generalizations thereof) of  ``hidden symmetries" of supergravities and M-theory \cite{Julia:1982gx,West:2001as,Damour:2002cu}.  While there is overwhelming evidence for the presence of these symmetries, the current results obtained so far remain incomplete because of the still somewhat mysterious role played by the ``dual graviton".  The clue for unlocking the present difficulties might precisely lie in a better grasp of the relationship between the graviton and its dual.

In \cite{Henneaux:2004jw}, two of the present authors presented a formulation of linearized gravity in four space-time dimensions that was manifestly invariant under ``duality rotations" in the space spanned by the graviton and its dual.  This work was further pursued in \cite{Bunster:2012km}, where it was shown in particular how the equations of motion following from the duality invariant action could be interpreted as twisted self-duality conditions on the curvature tensors of the graviton and its dual.   Duality invariance for linearized gravity was also considered from a different perspective in \cite{Deser:2004xt}.

The purpose of this paper is to extend the analysis of \cite{Henneaux:2004jw,Bunster:2012km} to linearized gravity in higher dimensions.  In that case the graviton and its dual are tensors of different types and so one cannot rotate them into one another.  However, one can still  write the equations of motion as twisted self-duality conditions on the curvature tensors: twisted self-duality remains although duality invariance is not present. Furthermore, as we shall show, one can also derive the twisted self-duality equations from a variational principle in which the graviton and its dual are on equal footing.  In fact, one may view the dual graviton as the canonically conjugate variable of the graviton, just as in the case of $p$-forms  \cite{Cremmer:1998px,Bunster:2011qp} or higher spins  \cite{Deser:2004xt}.  Particular important examples are $D=11$ (maximal supergravity/M-theory) and $D=5$ (first instance where the graviton and its dual are tensors of different types).

The crux to the twisted self-dual variational formulation relies on the Hamiltonian formulation and solving the constraints \cite{Deser:1976iy,Henneaux:2004jw,Bunster:2011qp,Bunster:2011aw}. This step introduces the prepotentials.  One may start either from the Hamiltonian formulation of the  action for the graviton, which is the Pauli-Fierz action, where the dynamical variables are the spatial components  of the metric  and their conjugate momenta, or from the  action for the dual graviton, where the dynamical variables are now the spatial components of the Curtright field and their conjugate momenta.  When expressed in terms of the prepotentials, the Pauli-Fierz and dual actions coincide.  One finds a remarkably symmetric situation, where the prepotential for the Pauli-Fierz field is at the same time the prepotential for the conjugate momentum to the Curtright field.  And similarly, the prepotential for the momentum conjugate of the  Pauli-Fierz field is at the same time the prepotential for the Curtright field.  In that sense, the Pauli-Fierz field and the Curtright field form a canonically conjugate pair since the Curtright field and the standard conjugate momentum to the Pauli-Fierz field just differ by linear redefinitions. 

Instead of expressing the action in terms of the prepotentials, one may keep the metric and trade its standard conjugate momentum  for the Curtright field.  This gives a first order action that involves symmetrically the graviton and its dual.  There remains constraints, however, and the action is non-local in space. 

For definiteness, we develop  the formalism in detail for the case $D=5$, where the dual graviton is a $(2,1)$ tensor.   As we indicate in the conclusions, the same construction applies to higher dimensions, but the formulas get more involved without new conceptual point.

Our paper is organized as follows.  In the next section (section \ref{DualForm}), we recall the dual formulation of linearized gravity in $D= d+1$ space-time dimensions, which involves a tensor of $(D-3,1)$ Young symmetry type. We give the action, the gauge symmetries and the invariants.  In section  \ref{LinTwisted}, we rewrite the linearized Einstein equations in arbitrary dimensions as twisted self-duality conditions and show in section \ref{Twisted} how to decompose them in space and time, in terms of  electric and magnetic components of the curvature tensors of the Pauli-Fierz field and its dual, which we explicitly define.  In section \ref{PrepPF}, we consider $D=5$ and introduce explicitly the prepotentials starting from the Pauli-Fierz action rewritten in Hamiltonian form.  The solution of  the momentum constraint introduces the prepotential for the conjugate momentum to the Pauli-Fierz field, which is a tensor of mixed $(2,2)$-Young symmetry type, while the solution of the Hamiltonian constraint introduces the prepotential for the Pauli-Fierz field, which is a tensor of mixed $(2,1)$-Young symmetry type (in addition to gauge prepotentials that drop out from the analysis).  These prepotentials are then shown in section \ref{PrepCurt} to emerge also from the Hamiltonian formulation of the Curtright action but their roles are now reversed.   The Pauli-Pierz action and the Curtright action are shown to coincide when written in terms of the prepotentials. The Poisson bracket structure is explored in section \ref{PBStructure} and the brackets between the curvatures are explicitly computed.   Section \ref{NoPrep} verifies  that the variational equations are the twisted self-dual conditions on the curvatures.  It also provides a formulation of the theory in which the variables are the spatial components of the Pauli-Fierz field and of its dual Curtright field. This formulation without prepotentials  is non-local in space (but local in time).  Finally section \ref{Conclusions} is devoted to comments and conclusions.

Four appendices support the analysis.   Appendix \ref{Weyl01} defines the analog of the Weyl tensor and the Cotton tensor for the Curtright field. These tensors are invariant under the analogs of the linearized Weyl rescalings of the Curtright field.  It is shown that in 4 dimensions, the Weyl tensor of the Curtright field vanishes but that the Cotton tensor is not zero.  We also consider the case of a spatial tensor of $(2,2)$ Young symmetry type as this is relevant for understanding some properties of the prepotentials.  Appendix \ref{AppDecomp} is devoted to the decomposition in space and time of the Riemann, Ricci and Weyl tensors.  The analog of the extrinsic curvature is defined for the Curtright field.  Further properties of the electric and magnetic fields associated with the graviton and its dual are also analyzed.  Appendix \ref{SHCCT} provides the explicit solution to the Hamiltonian constraints of the Curthright theory, which is somewhat technical. Finally, Appendix \ref{AppD} gives the inversion formulas for the prepotentials in terms of the original canonical variables. 

\section{Dual Formulation of Gravity}
\label{DualForm}
\setcounter{equation}{0}

\subsection{Sandard description of a spin-2 massless field: the Pauli-Fierz theory}
The standard description of a free massless spin-two particle involves a symmetric tensor $h_{\mu \nu} = h_{\nu \mu}$ (Young symmetry type $(1,1) \equiv \yng(2)$) subject to the linearized Einstein equations
\begin{equation}
R_{\lambda \rho} = 0. \label{LinearizedEinstein}
\end{equation}
Here, $R_{\lambda \rho}$ is the linearized Ricci tensor,
\begin{equation}
R_{\lambda \rho} = R_{\lambda \mu \rho \sigma} \eta^{\mu \sigma}  \label{Riemann0}
\end{equation}
where $R_{\lambda \mu \rho \sigma}$ is the linearized Riemann (``curvature") tensor,
\begin{equation}
R_{\lambda \mu \rho \sigma} = -\frac{1}{2} \left(\partial_\lambda \partial_\rho h_{\mu \sigma} -  \partial_\mu \partial_\rho h_{\lambda \sigma} - \partial_\lambda \partial_\sigma h_{\mu \rho} + \partial_\mu \partial_\sigma h_{\lambda \rho}\right).
\end{equation}

The Riemann tensor is of Young symmetry type
$$
(2,2) \equiv \yng(2,2)
$$
i.e., fulfills the identities 
\begin{equation}
R_{\lambda \mu \rho \sigma} =  R_{[ \lambda \mu ]\rho \sigma}, \; \; \; \; R_{\lambda \mu \rho \sigma} = R_{\lambda \mu [\rho \sigma ]}, \; \; \; \; \; R_{[\lambda \mu \rho] \sigma} = 0. \label{AlgebraicI}
\end{equation}
Here and in the sequel of this paper, brackets denote antisymmetrization of weight one, i.e., $F_{[\lambda \mu]}= \frac12 \left(F_{\lambda \mu} - F_{\mu \lambda} \right)$, so that, for instance, the condition $R_{\lambda \mu \rho \sigma} =  R_{[ \lambda \mu ]\rho \sigma}$ is equivalent to $R_{\lambda \mu \rho \sigma} = - R_{\mu \lambda \rho \sigma}$.  

In addition to (\ref{AlgebraicI}), the Riemann tensor also fulfills the Bianchi identity
\begin{equation}
\partial_{[\alpha_1} R_{\alpha_2 \alpha_2] \beta_1 \beta_2} = 0  \label{BianchiI}
\end{equation}
from which follows $R_{\alpha_2 \alpha_2[ \beta_1 \beta_2, \beta_3]} = 0$. Conversely, given a tensor $R_{\lambda \mu \rho \sigma}$ fulfilling the conditions (\ref{AlgebraicI}) and (\ref{BianchiI}), there is a tensor $h_{\lambda \mu}$ from which $R_{\lambda \mu \rho \sigma}$ derives as in (\ref{Riemann0}).  The tensor $h_{\lambda \mu}$ is determined up to a gauge transformation,
\begin{equation}
h_{\lambda \mu} \; \; \; \longrightarrow \; \; \;  h'_{\lambda \mu} = h_{\lambda \mu} + \partial_\lambda \xi_\mu + \partial_\mu \xi_\lambda 
\end{equation}
where $\xi_\mu$ is arbitrary.

The Riemann tensor and its derivatives give a complete set of gauge invariant objects, in the sense that any gauge invariant function is a function of the graviton field $h_{\lambda \mu}$ and its derivatives only through the Riemann tensor and its derivatives.  In particular, there is no gauge invariant function that depends only on $h_{\lambda \mu}$ and its first derivatives.  One must go to second derivatives to make gauge invariant objects, in contrast with the photon field and, more generally, $p$-form gauge fields.  In addition, a necessary and sufficient for the graviton field to be pure gauge is that the Riemann tensor vanishes \cite{Coho,DVH,XBNB}.

The action from which the equations of motion derive is the Pauli-Fierz action, i.e., the linearized Einstein-Hilbert action. It reads explicitly
\begin{equation}
S=\int d^{D}x \cal{L}
\end{equation}
where
\begin{equation}
{\cal{L}}=-\frac{1}{4}\left[\partial^{\rho}h^{\mu\nu}\partial_{\rho}h_{\mu\nu}-2\partial_{\mu}h^{\mu\nu}\partial_{\rho}h^{\rho}_{\ \nu}+2\partial^{\mu}h\partial^{\nu}h_{\mu\nu}-\partial^{\mu}h\partial_{\mu}h\right].
\end{equation}
The Lagrangian is quadratic in the first derivatives of $h_{\lambda \mu}$ and invariant under gauge transformations only up to a total derivative.

\subsection{The ``dual" graviton}

There exists a dual formulation of linearized gravity in terms of a tensor field with Young symmetry type
\begin{equation}
D-3 \hbox{ boxes} \left\{\yng(2,1,1,1,1,1,1,1) \right.
\end{equation}
In order to explain this point, we recall some background material.

\subsubsection{The Curtright field}
\label{CurtrightField}

In \cite{Curtright:1980yk}, the theory of generalized gauge fields described by higher rank tensors which are neither completely symmetric nor completely antisymmetric was initiated.   In particular, the simplest case of a tensor $T_{\alpha_1 \alpha_2 \beta}$ with mixed symmetry  $(2,1)$ corresponding to the Young tableau
\begin{equation}
\yng(2,1)
\end{equation} i.e.,
\begin{equation}
T_{\alpha_1 \alpha_2 \beta}= -T_{\alpha_2 \alpha_1 \beta}, \; \; \; \; T_{[\alpha_1 \alpha_2 \beta]} = 0
\end{equation}
was investigated in depth. The (free) action was shown to be
\begin{equation}
S [T_{\alpha_1 \alpha_2 \beta}] = -\frac16\int d^Dx \left[F_{\alpha_1 \alpha_2 \alpha_3 \beta} F^{\alpha_1 \alpha_2 \alpha_3 \beta} - 3 F_{\alpha_1 \alpha_2 \beta}^{\; \; \; \ \; \; \;\; \; \;  \beta} F^{\alpha_1 \alpha_2 \beta}_{\; \; \; \ \; \; \; \; \; \;  \beta} \right] \label{ActionCurt}
\end{equation}
where $F_{\alpha_1 \alpha_2 \alpha_3 \beta} = F_{[\alpha_1 \alpha_2 \alpha_3] \beta}$ is
\begin{equation}
F_{\alpha_1 \alpha_2 \alpha_3 \beta} = 3 \partial_{[\alpha_1} T_{\alpha_2 \alpha_3] \beta}.
\end{equation} 

The action (\ref{ActionCurt}) is invariant under the following gauge transformations of the ``Curtright field" $T_{\alpha_1 \alpha_2 \beta}$,
\begin{equation}
\delta T_{\alpha_1 \alpha_2 \beta} = 2 \partial_{[\alpha_1} \si_{\alpha_2] \beta} + 2 \partial_{[\alpha_1} \al_{\alpha_2] \beta} - 2 \partial_{\beta} \al_{\alpha_1 \alpha_2}\end{equation}
where $\si_{\alpha \beta}$ and $\al_{\alpha \beta}$ are symmetric and antisymmetric tensor fields, respectively,
\begin{equation}
\si_{\alpha \beta} = \si_{\beta \alpha}, \; \; \; \; \al_{\alpha \beta} = - \al_{\beta \alpha}.
\end{equation}  The field $F_{\alpha_1 \alpha_2 \alpha_3 \beta}$ is not gauge invariant but transforms  as
\begin{equation}
\delta F_{\alpha_1 \alpha_2 \alpha_3 \beta} =- 6  \partial_\beta \partial_{[\alpha_1} \al_{\alpha_2 \alpha_3]}
\end{equation}  

As a result, the Curtright Lagrangian is invariant only up to a total derivative. This situation is familiar in the context of the Chern-Simons action for a p-form.

In order to construct gauge invariant objects, one needs to take one more derivative, just as for the Pauli-Fierz field, whose linearized curvature contains two derivatives of the graviton field $h_{\mu \nu}$.  The ``curvature tensor" for the Curtright field is given by
\begin{equation}
E_{\alpha_1 \alpha_2 \alpha_3 \beta_1 \beta_2} =  2 F_{\alpha_1 \alpha_2 \alpha_3 [\beta_1, \beta_2] }
\end{equation}
and is easily checked to be gauge invariant and to have the Young tableau symmetry
\begin{equation}
(3,2) \equiv \yng(2,2,1)\; .
\end{equation} 
The curvature $E_{\alpha_1 \alpha_2 \alpha_3 \beta_1 \beta_2}$ enjoys the important property that it vanishes if and only if the Curtright field is pure gauge \cite{Coho}. Furthermore, it provides a complete set of gauge invariant objects, in the sense that any gauge invariant function of the Curtright field and its derivatives is a function of the curvature $E_{\alpha_1 \alpha_2 \alpha_3 \beta_1 \beta_2}$.  In analogy with the corresponding concepts of  the standard description of a spin-2 field, we call the tensor $E_{\alpha_1 \alpha_2 \alpha_3 \beta_1 \beta_2}$ the ``Riemann tensor" and associate with it the ``Ricci tensor" $E_{\alpha_1 \alpha_2 \beta} $ defined by 
\begin{equation}
E_{\alpha_1 \alpha_2 \beta} = E_{\alpha_1 \alpha_2 \gamma \beta_1}^{ \; \; \; \; \; \; \; \; \; \; \; \; \; \;  \gamma}
\end{equation}
as well as the ``vector curvature" $E_{\al}$,
\begin{equation}
E_{\alpha} = E_{\alpha  \gamma }^{ \; \; \; \; \;   \gamma}.
\end{equation}
One then introduces the ``Einstein tensor"  $G_{\alpha_1 \alpha_2 \beta}$ through
\begin{equation}
G_{\alpha_1 \alpha_2 \beta}= E_{\alpha_1 \alpha_2 \beta} + \frac{1}{2} (\eta_{\al_1 \beta} E_{\al_2} - \eta_{\al_2 \beta} E_{\al_1}).
\end{equation}
Both $E_{\alpha_1 \alpha_2 \beta}$ and $G_{\alpha_1 \alpha_2 \beta}$ are of $(2,1)$ Young symmetry type.  The Einstein tensor fulfills furthermore the ``contracted Bianchi identities"
\begin{eqnarray}
&&  \partial^{\alpha_2} G_{\alpha_1 \alpha_2 \beta} = 0 \\
&&  \partial^\beta G_{\alpha_1 \alpha_2 \beta} = 0
\end{eqnarray}
as a result of the Bianchi identities $\partial_{[\al_0} E_{\alpha_1 \alpha_2 \alpha_3] \beta_1 \beta_2}=0$ and $ E_{\al_1 \al_2 \al_3 [ \rho_1 \rho_2, \rho_3]}=0$ for the Riemann tensor.  One has also the doubly-contracted Bianchi identity
\be
\partial^\al E_\al = 0.
\ee

The field equations  that follow from the Curtright action are
\begin{equation}
G_{\alpha_1 \alpha_2 \beta} = 0 \label{EOMCurthright0}
\end{equation}
and equivalent to
\begin{equation}
E_{\alpha_1 \alpha_2 \beta} = 0. \label{EOMCurthright}
\end{equation}
 As shown in \cite{Curtright:1980yk}, the gauge invariances of the theory and the field equations enable one to go to a gauge in which the non-vanishing components of $T_{\alpha_1 \alpha_2 \beta}$ are transverse and traceless, and obey furthermore the Laplace equation.   The relevant irreducible representation of the little group $SO(D-2)$ is described by the same Young tensor (with the additional trace conditions) and yields
\begin{equation}
N_0(D) = \frac13 D (D-2)(D-4)
\end{equation}
physical degrees of freedom (``helicity states").   

Observe that in $D=4$ dimensions, where the little group is $SO(2)$, a $(2,1)$-tensor described by the Curtright action carries no degree of freedom since $N_0(D) = 0$.  One way to understand this fact is that the  $SO(2)$-irreducible conditions $T_{ijk} = T_{[ij]k}$, $T_{[ijk]} = 0$ and $T_{ij}^{\; \; \;   j}=0$ imply $T_{ijk} =0$.  Another way to see the same thing is to observe that in $D=4$ dimensions, the curvature tensor of the Curtright field is completely determined by the Ricci tensor through
\begin{eqnarray}
\hspace{-1cm} E_{\al \beta \gamma \rho \si} &=&  
 -  \Big[ ( \eta_{\al \rho} S_{\beta \gamma \si}  - \eta _{\al \si} S_{\beta \gamma \rho})   
\nonumber \\ && \hspace{1.5cm} + (\eta_{\beta \rho} S_{ \gamma \al \si} - \eta _{\beta \si} S_{ \gamma \al \rho}) + (\eta_{\gamma \rho} S_{\al \beta \si} - \eta _{\gamma \si} S_{\al \beta  \rho}) \Big]  
\end{eqnarray}
where the ``Schouten tensor" $S_{\al \bet \rho}$ is given by
$$
S_{\al \bet \rho} = E_{\al \bet \rho} + \frac{1}{4}(\eta_{\al \rho} E_\beta - \eta_{\beta \rho} E_\al).
$$
It follows that $E_{\alpha_1 \alpha_2 \alpha_3 \beta_1 \beta_2}$ vanishes by the field equations and the field is pure gauge.

One can introduce for the Curtright field the useful concepts of Weyl and Cotton tensors.  This is done in the appendix \ref{Weyl01}.  Note that we are using the same letters $G$, $S$ etc for the Einstein, Schouten etc tensors of the Pauli-Fierz and Curtright fields.  The number of indices being distinct tells the difference and there is no risk of confusion.  When we want to emphasize the difference, however, we shall complete the notation by adding the argument ``$h$" or  ``$T$" between brackets, e.g., $G_{\al \beta}[h]$, $G_{\al \beta \ga} [T]$ etc.

\subsubsection{Generalisation}

The above construction can readily be extended to tensor fields $T_{\alpha_1 \alpha_2 \cdots \alpha_{k} \beta}$  with mixed symmetry 
\begin{equation}
T_{\alpha_1 \alpha_2 \cdots \alpha_{k} \beta}= T_{[\alpha_1 \alpha_2 \cdots \alpha_{k}] \beta}, \; \; \; \; T_{[\alpha_1 \alpha_2 \cdots \alpha_{k} \beta]}= 0
\end{equation}
 corresponding to the Young tableau
\begin{equation}
k \hbox{ boxes} \left\{\yng(2,1,1,1,1,1,1,1) \right.
\end{equation}
The action and the gauge invariance have been given in \cite{Aulakh:1986cb,Labastida:1986gy}, where it was also shown that the physical degrees of freedom are massless and carried by  the transverse and traceless components $T_{i_1 i_2 \cdots i_{k} j}$ of  $T_{\alpha_1 \alpha_2 \cdots \alpha_{k} \beta}$ ($i_m, j = 1, \cdots, D-2$), which transform in the $SO(D-2)$ representation described by the same Young tableau.

\subsubsection{Gravitational Duality}

When $k = D-3$, the irreducible representations of the little group $SO(D-2)$ described by the Young tableaux
\begin{equation}
\yng(2)
\end{equation}
and 
\begin{equation}
D-3 \hbox{ boxes} \left\{\yng(2,1,1,1,1,1,1,1) \right.  \label{DualGrav}
\end{equation}
are equivalent, as can be seen through the relation $$h_{ij} = \frac12 \epsilon_{im_1m_2 \cdots m_{D-3}} T^{m_1m_2 \cdots m_{D-3}}_{\; \; \; \; \; \; \; \; \; \; \; \; \; \; \; \; \; \; \; \; \; \; \; \; j},$$ using trace and antisymmetry conditions. Therefore, the Pauli-Fierz action and the actions of \cite{Curtright:1980yk} ($D=5$) and \cite{Aulakh:1986cb,Labastida:1986gy} ($D \geq 6$) provide equivalent, dual descriptions of linearized gravity.  The field (\ref{DualGrav}) is called the ``dual graviton". For $D=4$,  the dual graviton is a symmetric tensor like the graviton itself  since $D-3 = 1$ and the dual description of gravity is  given by the Pauli-Fierz action.

The equivalence of the (on-shell) physical modes of the gauge-invariant theory based on a  tensor field with the mixed $(D-3,1)$ symmetry with the (on-shell) physical modes of linearized gravity was observed for $D=5$ in \cite{Curtright:1980yk} and in higher dimensions in \cite{Hull:2000zn}.  

{}This equivalence can be established from a different perspective.   In the context of the conjectured hidden symmetry $E_{11}$ of $M$-theory, a remarkable property  was discovered in \cite{West:2001as}, namely, that   $E_{11}$  implies  the existence of a tensor field of mixed $(8,1)$ symmetry type $T_{\alpha_1 \alpha_2 \cdots \alpha_{8} \beta}$ ($D-3=8$).  This follows from  the decomposition of the $E_{11}$ adjoint representation in terms of space-time tensors. Such a tensor also occurs in the $E_{10}$ formulation \cite{Damour:2002cu}. It was furthermore indicated in \cite{West:2001as} that the tensor $T_{\alpha_1 \alpha_2 \cdots \alpha_{8} \beta}$ should be the dual to the graviton. This  insight was obtained by starting from Einstein's theory  in $D$ dimensions in the first-order formulation in terms of the vielbein and a mixed tensor field $Y_{\al_1\ldots  \al_{D-2}\beta}$ with the Young symmetry $(D-2,1)$, related to the standard spin connection through algebraic redefinitions.  This connection-like tensor was then shown to be equal, at the linearized level, to the exterior derivative of a field $T_{\al_1\ldots \al_{D-3}, \beta}$ with the required symmetry type $(D-3,1)$.         This followed from the vielbein field equations which appeared, at the linearized level, as a Lagrange multipliers for the constraint $\partial_{[\al_0}Y_{\al_1\ldots  \al_{D-2}]\beta}=0$ \cite{West:2001as}. That the resulting ``potential" $T_{\al_1\ldots \al_{D-3}, \beta}$ that solved the constraint was in fact described by the action of \cite{Curtright:1980yk,Aulakh:1986cb,Labastida:1986gy}, with the correct gauge symmetries, was then established in \cite{Boulanger:2003vs}, extending the on-shell equivalence derived in the light cone to full off-shell equivalence, i.e., equivalence at the level of the action.

Finally, we point out that explicit on-shell expressions for the dual graviton field in terms of the original graviton field have been given in \cite{Bakas:2009da} for four dimensions and can be generalized to higher dimensions.


\section{Linearized Einstein equations as twisted self-duality equations for the curvatures}
 \setcounter{equation}{0}
 \label{LinTwisted}

We shall from now on assume $D=5$.  This is done only for the sake of keeping formulas simple and implies no conceptual restriction. The extension to the general case is outlined in the conclusions. As we have just seen, the graviton and its dual are then described by tensor fields with the respective Young symmetries
$
\yng(2) $ and $
\yng(2,1) $.
One can interpret the gravitational duality equations in terms of the curvatures as follows \cite{Hull:2001iu}. 

The Einstein equations  $R_{\mu \nu} = 0$ for the Riemann tensor $R_{\mu \nu \al \beta}[h]$ imply that the dual Riemann tensor $E_{\beta_1 \beta_2 \beta_3 \rho_1 \rho_2}$, defined by
\begin{eqnarray}
E_{\beta_1 \beta_2 \beta_3 \rho_1 \rho_2} &=&  \frac{1}{2!} \epsilon_{ \beta_1 \beta_2 \beta_3 \al_1 \al_2 } R^{\al_1 \al_2}_{\; \; \; \; \; \; \; \; \rho_1 \rho_2} \nonumber \\
R_{\al_1 \al_2 \rho_1 \rho_2} &=& -\frac{1}{3!}\epsilon_{\al_1 \al_2 \beta_1 \beta_2 \beta_3}  E^{\beta_1 \beta_2 \beta_3}_{\; \; \; \; \; \; \; \; \; \; \; \;  \rho_1 \rho_2}  \nonumber 
\end{eqnarray}
is of Young symmetry type 
$$\yng(2,2,1)\; .$$
Our conventions are $\epsilon_{01234} = 1 = - \epsilon^{01234}$, so that in particular $\epsilon_{0ijk\ell} =  \epsilon_{ijk\ell}$.
Furthermore, (i) the tensor  $E_{\beta_1 \beta_2 \beta_3 \rho_1 \rho_2}$ obeys the differential identities $ \partial_{[\beta_0}E_{\beta_1 \beta_2 \beta_3 ]\rho_1 \rho_2}  = 0$,  $E_{\beta_1 \beta_2 \beta_3 [\rho_1 \rho_2 ,\rho_3]}= 0$  that guarantee the existence of a tensor $T_{\al \beta \mu}$ such that 
$$ E_{\beta_1 \beta_2 \beta_3 \rho_1 \rho_2} = E_{\beta_1 \beta_2 \beta_3 \rho_1 \rho_2}[T] $$
as in subsection \ref{CurtrightField}.; and (ii) the field equations (\ref{EOMCurthright}) for the dual tensor $T_{\al \beta \mu}$ are satisfied. 

Conversely, one may reformulate the gravitational field equations as twisted self-duality equations as follows.  Let $h_{\mu \nu}$ and $T_{\al \beta \mu}$ be tensor fields of respective Young symmetry types $\yng(2)$ and $\yng(2,1)$, and let $R_{\al_1 \al_2 \rho_1 \rho_2}[h]$ and $E_{\beta_1 \beta_2 \beta_3 \rho_1 \rho_2}[T]$ be the corresponding gauge-invariant curvatures.  The  ``twisted self-duality conditions", which express that $E$ is the dual of $R$ (we drop indices)
$$
R =- \; \; \!^*E, \; \; \; E = \;    \!^*R,
$$
or, in matrix notations,
\begin{equation}
{\mathfrak R} =  {\mathcal S}  \, ^*\hspace{-.02cm}{\mathfrak R}, \label{122}
\end{equation}
with
\begin{equation}
{\mathfrak R} = 
 \begin{pmatrix} R\\ E\\ \end{pmatrix}, \; \; \; {\mathcal S}  = \begin{pmatrix} 0&-1 \\ 1 & 0 \end{pmatrix}  ,\label{123}
\end{equation}
imply that $h_{\mu \nu}$ and $T_{\al \beta \mu}$ are both solutions of the linearized Einstein equations and the Curtright equations,
$$
R_{\mu \nu} = 0, \; \; \;  E_{\mu \nu \alpha} = 0.
$$
This is because, as we have seen, the cyclic identity for $E$ (respectively, for $R$) implies that the Ricci tensor of $h_{\al \beta}$ (respectively, of $T_{\al \beta \ga}$) vanishes.

The equations (\ref{123}) are called twisted self-duality conditions for linearized gravity because if one views the curvature ${\mathfrak R}$ as a single object, then the conditions (\ref{123}) express that this object is self-dual up to a twist, given by the matrix ${\mathcal S}$.  The twisted self-duality equations put  the graviton and its dual  on an identical footing. 

The twisted self-duality equations relate gauge-invariant objects.  They share a great similarity with the twisted self-duality formulation of $p$-form field equations \cite{Cremmer:1998px}.   However, there is one important difference.  Because the gauge-invariant curvatures for the spin-2 field contains second order derivatives, the twisted self-duality conditions are second order partial differential equations for $h_{\mu \nu}$ and $T_{\mu \nu \alpha}$, while these conditions are first-order differential equations for the  potentials in the case of $p$-forms.

In the next section, we examine more closely the twisted self-duality conditions for gravity and show that there is a subset of them which contains only first-order time derivatives and which is complete, in the sense that the entire set of twisted self-duality conditions follows from it.   Although of first-order in the time derivatives, these equations contain second order spatial derivatives.


\section{$4+1$-Form of the Twisted Self-Duality Conditions}
\label{Twisted}
\setcounter{equation}{0}
\subsection{Constraint equations - Dynamical equations}
Just as the Einstein equations split into constraint equations,
$$
G_{00} \equiv R_{00} + \frac12 \, ^{(5)}\!R= 0, \; \; \; G_{0i} \equiv R_{0i} = 0
$$
and dynamical equations,
$$
^{(5)} \! G_{ij} \equiv  ^{(5)} \! R_{ij}- \frac12 \, \delta_{ij} \, ^{(5)}\!R = 0 \, ,
$$
so do the equations of motion for the Curtright field.  Here and in what follows, we  affect the spacetime objects with an index $^{(5)}$ when a possible confusion can arise.  So, for instance,  $^{(5)} \! R_{ij}$ denotes the $(i,j)$ component of the Ricci tensor of the spacetime metric while $R_{ij}$ (sometimes also written $^{(4)} \! R_{ij}$ to emphasize the difference) denotes the $(i,j)$ component of the Ricci tensor of the spatial metric.  As Eq. (\ref{DefR}) below indicates, $^{(5)} \! R_{ij} \not= \, ^{(4)} \! R_{ij}$.

More precisely, the $0i0$ and $0ij$ components of the variational equations of motion do not contain second order time derivatives of the field and are therefore constraints,
$$
G_{0i0} = 0, \; \; \; G_{0ij} = 0 \; \; (\Rightarrow G_{ij0} = G_{0ji} - G_{0ij} = 0)
$$
while the $ijk$ components are the dynamical equations,
$$
^{(5)} \! G_{ijk} = 0.
$$
We leave the verification of this direct property to the reader. Useful expressions for the space and time decomposition of the Einstein/Ricci tensors worked out in  Appendix \ref{AppDecomp} are
\begin{eqnarray}
&& G_{00} =  \frac12 \; ^{(4)} \! R\\
&& R_{0i} = - \partial^m (K_{im} - \delta_{im} K)\\
&& ^{(5)} \! R_{ij} = - \partial_0 K_{ij} + \frac12 \, \partial_i \partial_j h_{00} +  \, ^{(4)} \!R_{ij}  \label{DefR}\\
&& G_{0i0} = - \frac{1}{2} \, ^{(4)} \! E_{ik}^{\; \; \; \;  k}\\
&& E_{0ij} = \partial^k V_{ikj} - \partial_j V_{ik}^{\; \; \; \; k} \\
 && ^{(5)} \! E_{ijk} =   - \partial_0 V_{ijk} - \partial_k (\partial_i T_{0j0} - \partial_j T_{0i0}) + \, ^{(4)}E_{ijk} \label{5E4E}
\end{eqnarray}
where $K_{ij}$ is the extrinsic curvature and $V_{ijk}$ the invariant velocity for the Curtright field introduced in  Appendix \ref{AppDecomp}.  Explicitly,
$$
K_{ij} = - \frac12 (\dot{g}_{ij} - g_{0i},_{j}- g_{0j},_{i})
$$
and
$$
V_{ijk} = \dot{T}_{ijk} + \partial_i T_{j0k} - \partial_j T_{i0k} - \partial_k T_{ij0}.
$$
These variables are ``invariant velocities" in the sense of Dirac \cite{Dirac}.  Namely, their gauge transformations on any spacelike hypersurface depend only on the gauge parameters on that hypersurface, and not on their derivatives off that hypersurface.

It will be convenient in the sequel to use the following equivalent set of equations of motion:
\begin{itemize}
\item For the Einstein equations: $G_{00} = 0, R_{0i} = 0$ (constraints), $^{(5)} \! R_{ij} = 0$ (dynamical equations)
\item For the Curtright equations: $G_{0i0} = 0, E_{0ij} = 0$ (constraints), $^{(5)} \! E_{ijk} = 0$ (dynamical equations)
\end{itemize}

\subsection{Electric and Magnetic Fields}

It is useful to introduce the ``electric" and ``magnetic" fields built out of the Riemann tensor.  These are defined as follows:
\begin{itemize}
\item For the standard graviton: 
\begin{eqnarray}
{\mathcal E}_{ijrs}[h] & = &\frac{1}{4} \epsilon_{ijmn} R^{mnpq} \epsilon_{pqrs} \\
& =  &  R_{ijrs} - \delta_{ir} R_{js}+  \delta_{is} R_{jr} + \delta_{jr} R_{is} -  \delta_{js}  R_{ir}  \nonumber \\ && + \frac12 (\delta_{ir} \delta_{js} - \delta_{is} \delta_{jr})R
\end{eqnarray} 
and
\begin{equation}
{\mathcal B}_{rsi} [h] =  \frac12 \epsilon_{rsmn}R_{0i}^{\; \; \; \;   mn}
\end{equation}
(note that $^{(5)} \! R_{ijrs}[h] = R_{ijrs}[h]$).
\item For the dual graviton
\begin{equation} {\mathcal E}_{ijr} [T]=  G_{ijr}, \;  \; \; \; {\mathcal B}_{ijrs} [T] =  \frac{1}{2}\epsilon_{rsmn}E_{0ij}^{\; \; \; \; \;  mn}
\end{equation}
where $G_{ijr}$ is the spatial Einstein tensor constructed out of the spatial tensor $T_{ijm}$, $G_{ijr} \equiv ^{(4)} \! G_{ijr}$ (note that $^{(5)} \! E_{ijkrs}[h] = E_{ijkrs}[h]$ and that in four dimensions, $E_{ijkrs}[h]$ is compeletely determined by $E_{ijr}[h]$).
\end{itemize}
The electric fields depend on the second spatial derivatives of the spatial components $h_{ij}$ and $T_{ijr}$ of the Pauli-Fierz or Curthright fields, respectively, and involve no time derivative.  
The magnetic fields involve only the spatial gradients of the invariant velocities $K_{ij}$, $V_{ijr}$, respectively.  They do not involve neither $h_{00}$ nor $T_{0i0}$. Further properties of the electric and magnetic fileds are discussed in Appendix \ref{AppDecomp}.  These can be summarized as:
\begin{itemize} 
\item The electric fields transform identically in the $\yng(2,2)$- and $\yng(2,1)$-representations, i.e., their irreducible components not in these representations identically vanish by definition.
\item The conditions ${\mathcal B}_{[rsi]} [h]=0 $ (respectively ${\mathcal B}_{[ijr]s} [T] =0$), which express that the magnetic field transform in the $\yng(2,1)$-representation (respectively, the $\yng(2,2)$-representation), are equivalent to the constraint equations $R_{0i}=0$ (respectively, $E_{0ij}= 0$)
\item  The electric and magnetic fields are identically transverse as a consequence of the Bianchi identities,
\begin{eqnarray}
&& \partial^m {\mathcal E}_{mnrs}[h] = 0 , \; \; \;  \partial^m {\mathcal E}_{mn} [h] = 0 , \nonumber \\
&& \partial^m {\mathcal E}_{mnr}[T] = 0 , \nonumber \\
&& \partial^r {\mathcal B}_{mnrs}[T] = 0 , \; \; \;  \partial^n {\mathcal B}_{mn} [T] = 0 , \nonumber \\
&& \partial^m {\mathcal B}_{mnr}[h] = 0. \nonumber
\end{eqnarray}
\item The electric field ${\mathcal E}_{ijrs} [h]$ (respectively ${\mathcal E}_{ijr} [T]$) has a vanishing double-trace (respectively, vanishing traces) as a result of the constraint equations $G_{00}[h]=0$ (respectively $G_{0i0}[T]=0$) while the magnetic fields fulfill these conditions identically,
\begin{eqnarray}
&&  {\mathcal E}_{mnrs}[h] \delta^{ns} \delta^{mr} = 0 ,  \nonumber \\
&&  {\mathcal E}_{mnr}[T] \delta^{nr} = 0  , \nonumber \\
&&  {\mathcal B}_{mnrs}[T] \delta^{ns} \delta^{mr} = 0 , \nonumber \\
&&  {\mathcal B}_{mnr}[h] \delta^{nr} = 0. \nonumber
\end{eqnarray}
Conversely, the above first two equations imply the constraints $G_{00}[h]=0$ and  $G_{0i0}[T]=0$.
\end{itemize}

\vspace{.3cm}
 
It follows from the twisted self-duality conditions that,
\begin{eqnarray}
&& {\mathcal B}_{ijr}[h] = - {\mathcal E}_{ijr}[T] , \nonumber \\
&& {\mathcal B}_{ijrs}[T] =  {\mathcal E}_{ijrs}[h] ,\nonumber 
\end{eqnarray}
or in matrix notation,
\begin{equation}
 \begin{pmatrix} \mathcal{B}_{ijr}[h]\\ \mathcal{B}_{ijrs}[T]\\ \end{pmatrix} = {\mathcal S}  \, \begin{pmatrix} \mathcal{E}_{ijrs}[h]\\ \mathcal{E}_{ijr}[T]\\ \end{pmatrix} .\label{EBDuality}
\end{equation}

\subsection{More on the Twisted Self-duality Conditions}

The twisted self-duality conditions (\ref{EBDuality}) are the purely spatial components 
$$ R_{ijrs} = \frac12 \epsilon_{ij0mn} E^{0mn}_{\; \; \; \; \; \; \; \; rs} , \; \; \; \; \; E_{ijkrs} = - \epsilon_{ijk0m} R^{0m}_{\; \; \; \; \; rs},
$$
of the covariant twisted self-duality conditions (\ref{122}), 
and so, form only the subset of these equations that do not involve the components $R_{0i0r}$ and $E_{0ij0r}$ with two time indices of the curvatures.
We claim that the twisted self-duality conditions (\ref{EBDuality}) are nevertheless completely equivalent to (\ref{122}) and in particular imply all of the Einstein equations.  

To establish the claim, we proceed in two steps.

\subsection{Step 1: Constraint Equations}

The constraint equations follow from (\ref{EBDuality}).  

  \vspace{.1cm} 
  
  \noindent
  {\bf Proof:} It follows from the twisted self-duality relations (\ref{EBDuality}) that the magnetic fields ${\mathcal B}_{ijr}[h]$ and ${\mathcal B}_{ijrs}[T]$ fulfill ${\mathcal B}_{[ijr]}[h]=0$ and ${\mathcal B}_{[ijr]s}[T]=0$ since the corresponding electric fields do identically.  This implies  $R_{0i}[h] = 0$, $E_{0ij}[T] = 0$.

In a similar way, it follows from the twisted self-duality relations (\ref{EBDuality}) that the electric field ${\mathcal E}_{ijrs}[h]$ is double-traceless and  that ${\mathcal E}_{ijr}[T]$ is traceless, since the corresponding magnetic fields enjoy these properties identicallly.  This implies that the spatial curvatures  $R[h]$ and $ E_i[T]$ both vanish. For the linearized theory these are the constraints $G_{00}[h] = 0 = G_{0i0}[T]$ for $h_{\mu \nu}$ and $T_{\mu \nu \rho}$.

Once the constraints are established, one can verify that the Bianchi identities  imply the relations
\begin{equation}
\partial_0 {\mathcal E}_{ijkl}[h] = \frac12 \epsilon_{ijmn} (\partial^m {\mathcal B}_{kl}^{\; \; \; \; n}[h] -\partial^n {\mathcal B}_{kl}^{\; \; \; \; m}[h]) \label{BianchiEB0}
\end{equation}
and
\begin{equation}
\partial_0 {\mathcal E}_{rsi} [T] = \frac12 \partial^k {\mathcal B}_{rs}^{\; \; \; \; mn}[T] \epsilon_{ikmn} ,\label{BianchiEB1}
\end{equation}
which will be usedul later.  It is in the last relation that we have used the constraint equations.

\subsection{Step 2: Dynamical Equations}

The next step is to establish the dynamical Einstein equations.  This is a bit harder because these involve two time derivatives of the metric, so that one needs to differentiate the self-duality conditions with respect to time.  But this will lead to third order differential equations and so, one can only hope to get the dynamical Einstein equations differentiated once, but without loss of information.  This turns out to be the case.  This is in sharp contrast with the electromagnetic situation, where the twisted self-duality conditions are first-order differential equations, while the Maxwell equations are of second order, so that by differentiating once the twisted self-duality condition, one can derive the Maxwell equations in their standard form.

\vspace{.2cm}
These general considerations being stated, we now turn to the proof that the dynamical Einstein equations also follow from (\ref{EBDuality}).  

\vspace{.1cm}

\noindent

{\bf Proof:}  We compute  $\partial_0 R_{i0mj}[h]$ in two different ways.  On the one hand, it is equal to
\begin{equation} \partial_0 R_{i0mj}[h] = -\partial_m R_{i0j0}[h] + \partial_j R_{i0m0}[h], \label{TD1} \end{equation}
by the Bianchi identity.
On the other hand, $R_{i0mj}[h] = - \frac12 \epsilon_{mj}^{\; \; \; \; \; rs} {\mathcal B}_{rsi}[h]$ and so,
\begin{eqnarray}
\partial_0 R_{i0mj}[h]  &=& - \frac12 \epsilon_{mj}^{\; \; \; \; \; rs} \partial_0 {\mathcal B}_{rsi}[h] ,\nonumber \\
&=&   \frac12 \epsilon_{mj}^{\; \; \; \; \; rs} \partial_0 {\mathcal E}_{rsi}[T] \label{TD2a}
\end{eqnarray}
by (\ref{EBDuality}).  Using successively (\ref{BianchiEB1}), the twisted self-duality condition again, the transverseness of the electric field and the constraint equation $R=0$,  this becomes 
\begin{eqnarray}
\partial_0 R_{i0mj}[h]  &=& - \frac14 \epsilon_{mj}^{\; \; \; \; \; rs} \epsilon_{ik}^{\; \; \; \; \; pq}\partial^k {\mathcal E}_{rspq}[h] ,\nonumber \\
&=&  \partial_m {\mathcal E}_{ij}[h] - \partial_j {\mathcal E}_{im}[h] .\label{TD2b}
\end{eqnarray}
Comparing (\ref{TD1}) with (\ref{TD2b}) gives,
$$
\partial_m \left( R_{i0j0}[h] + {\mathcal E}_{ij}[h]\right)- \partial_j \left(R_{i0m0}[h] + {\mathcal E}_{im}[h] \right)= 0.
$$
Taking into account the definition of  ${\mathcal E}_{ij}[h]$, this is just
\begin{equation}
-\partial_m  \! ^{(5)} \!R_{ij}[h] + \partial_j \! ^{(5)} \! R_{im}[h] = 0. \label{RotR}
\end{equation}

This equation is a bit misleading at first sight, since $\! ^{(5)} \!R_{ij}[h]$ contains $h_{00}$ while neither the electric field nor the magnetic field does, and we started with relations that involved only the electric and the magnetic fields.  But if one explicitly
plugs the expression (\ref{DefR}) in (\ref{RotR}), one sees that $h_{00}$ does drop out from (\ref{RotR}), as it should.  
 
To analyze the implications of the equation (\ref{RotR}), it is easier to write it in terms only of $h_{ij}$ and $h_{0i}$ as
\begin{equation}
\partial_m (\partial_0 K_{ij} - R_{ij}) - \partial_j \partial_0( K_{im}-R_{im}) = 0. \label{RotPartialK}
\end{equation}
 This equation implies, using the fact that  $\partial_0 K_{ij} $ and $R_{ij}$ are symmetric in $(i,j)$
$$
\partial_0 K_{ij} - R_{ij}= \partial_i \partial_j \Phi
$$
for some function  $\Phi$.  Choosing the function $h_{00}$ (which is an arbitrary gauge function not occurring in the original equations (\ref{EBDuality})) to be equal to $2 \Phi$ yields
$$
\! ^{(5)} \!R_{ij}[h] = 0.
$$
These are the dynamical Einstein equations for $h_{\mu \nu}$.  

Similarly, expressing the time derivatives of $E_{ij0kl}$ in two different ways yields the dynamical equations for the Curtright field.  The explicit steps are: one the one hand, the Bianchi identities imply 
\begin{equation}
\partial_0 E_{ij0kl} [T] = \partial_l E_{ij0k0} [T] - \partial_k E_{ij0l0}[T]. \label{BICurt1}
\end{equation}
On the other hand, \begin{eqnarray}
 \partial_0 E_{ij0kl} [T] &=& \frac12 \partial_0 {\mathcal B}_{ijrs}[T] \epsilon_{kl}^{\; \; \; \; \; rs} \nonumber \\
 &=&- \frac12 \partial_0 {\mathcal E}_{ijrs}[h] \epsilon_{kl}^{\; \; \; \; \; rs} \nonumber \\
&=& - \frac12  \epsilon_{ijmn} \epsilon_{kl}^{\; \; \; \; \; rs} \partial^m {\mathcal B}_{rs}^{\; \; \; \;  n} [h]\nonumber \\
&=&- \frac12  \epsilon_{ijmn} \epsilon_{kl}^{\; \; \; \; \; rs} \partial^m {\mathcal E}_{rs}^{\; \; \; \;  n} [T]\nonumber \\
&=& - \partial_k {\mathcal E}_{ijl}[T] +  \partial_l {\mathcal E}_{ijk}[T] \label{TimeDer2}
 \end{eqnarray}
 where we have successively used the definition of the magnetic field $ {\mathcal B}_{ijrs}[T]$, the twisted self-duality condition, the equation (\ref{BianchiEB0}), the twisted self-duality condition again, and the fact that the electric field is transverse and traceless.
 
 Comparing (\ref{BICurt1}) with (\ref{TimeDer2}) yields
 $$  \partial_l (E_{ij0k0} [T] - {\mathcal E}_{ijk}[T] ) - \partial_k (E_{ij0l0}[T]- {\mathcal E}_{ijl}[T])  = 0,$$
 i.e., 
 \begin{equation}
 \partial_l ^{(5)} \! E_{ijk} [T] - \partial_k ^{(5)} \! E_{ijl} [T]  = 0. \label{RotT}
 \end{equation}
 These equations are discussed in the same way as the equations (\ref{RotR}) for the Pauli-Fierz theory.  One easily verifies that the components $T_{0i0}$ of the Curthright field, which occur in $^{(5)} \! E_{ijk} [T]$,  drop out from the curl (\ref{RotT}), and that the equations(\ref{RotT})  imply through integration the dynamical Curtright equations
$$
 ^{(5)} \! E_{ijk} [T] = 0
$$
 by an appropriate adjustment of these gauge functions $T_{0i0}$. (From (\ref{RotT}) one gets first$^{(5)} \! E_{ijk} [T] = \partial_k \mu_{ij}$ for some $\mu_{ij} = - \mu_{ji}$.  The condition $^{(5)} \! E_{[ijk]} [T]  = 0$ implies then $\partial_{[k }\mu_{ij]} = 0$, i.e., $\mu_{ij} = \partial_i \lambda_j - \partial_j \lambda_i$ for some $\lambda_i$ with can be absorbed in a redefinition of $T_{0i0}$, see (\ref{5E4E}).)
 
We thus conclude that  the spatial twisted self-duality conditions  (\ref{EBDuality}) imply all of Einstein and Curtright equations.

\subsection{Search for a variational principle}
The twisted self-duality conditions put the graviton (Pauli-Fierz) field $h_{\mu \nu}$ and the dual graviton (Curtright) field $T_{\mu \nu \rho}$ on an equal footing.  The central goal of this paper is to derive a variational principle that keeps this democratic treatment of the graviton field and its dual. 

Neither the Pauli-Fierz second-order action principle nor the Curtright second order action principle treat the graviton and its dual on the same footing since they involve only one field, either the graviton or its dual, but not both simultaneously.

The ``mother action" considered in \cite{West:2001as,Boulanger:2003vs}, which enables one to go from one picture to its dual, does not provide the answer to the question because it involves the graviton field (in the frame formulation) and another field which is equal on-shell to the curl of the dual field through a constraint.  So the dual field does not enter  this action principle on the same footing as the graviton, since one is a fundamental field to be varied in the action principle while the other appears as a derived concept.  [And if one eliminates the constraint, one looses the original graviton field and gets the Curtright action.]  The dual ``father action" would have the roles of the graviton and its dual reversed and so would suffer from the same drawback.   The action proposed in the interesting work \cite{Boulanger:2008nd} shares similar features. 

It turns out that a variational principle does fulfills this condition of putting the graviton field and its dual on exactly the same footing. It is given below and is in fact just the Pauli-Fierz action in Hamiltonian form rewritten in terms of the appropriate variables. It yields the spatial part of the twisted self-duality conditions, which have been shown to form a complete set of equations. A striking feature is that it does not preserve manifest space-time covariance, a feature shared also by the duality-symmetric actions principles for $p$-form gauge fields \cite{ScSe,Bunster:2011qp}.  This is not necessarily a drawback as one might argue that space-time is a derived concept and that space-time covariance might follow from duality symmetry \cite{Bunster:2012hm}.


\section{Prepotentials - Starting from the Pauli-Fierz Action}
\label{PrepPF}
\setcounter{equation}{0}

As for $p$-form gauge fields \cite{Bunster:2011qp} and linearized gravity in four dimensions \cite{Henneaux:2004jw}, the requested duality-symmetric action where both the Pauli-Fierz and the Curtright fields appear on an equal footing is obtained by going to the Hamiltonian formalism and solving the constraints.

\subsection{Hamiltonian formulation}

The expression of the momenta $\pi^{mn}$ conjugate to $h_{mn}$ in terms of  the extrinsic curvature $K_{mn}$  is
$$
\pi^{mn} = - K^{mn} + K \delta^{mn}
$$
and its inverse reads
$$
K_{mn} = - \pi_{mn} + \frac{\pi}{D-2}  \delta_{mn},
$$
yielding the action in Hamiltonian form
\begin{equation}
S=\int dt d^{D}x\left[\pi_{mn}\dot{h}^{mn}-{\cal{H}}- nC -n_m C^m \right]
\end{equation}
where ${\cal{H}}$ is the Hamiltonian:
\begin{eqnarray}
{\cal{H}}&=& \pi_{mn}\pi^{mn}-\frac{\pi^{2}}{D-2}+\frac{1}{4}\partial_{r}h_{mn}\partial^{r}h^{mn}-\frac{1}{2}\partial_{m}h^{mn}\partial_{r}h^{r}_{\ n}+\nonumber \\
&& \hspace{2cm} + \frac{1}{2}\partial^{m}h\partial^{n}h_{mn}-\frac{1}{4}\partial_{m}h\partial^{m}h, \nonumber
\end{eqnarray}
i.e, for $D=5$,
\begin{eqnarray}
{\cal{H}}&=& \pi_{mn}\pi^{mn}-\frac{\pi^{2}}{3}+\frac{1}{4}\partial_{r}h_{mn}\partial^{r}h^{mn}-\frac{1}{2}\partial_{m}h^{mn}\partial_{r}h^{r}_{\ n}+\nonumber \\
&& \hspace{2cm} + \frac{1}{2}\partial^{m}h\partial^{n}h_{mn}-\frac{1}{4}\partial_{m}h\partial^{m}h.
\end{eqnarray}
The components $h_{0m}\equiv n_m$ and $h_{00}\equiv 2 n$ only appear linearly and multiplied by terms with no time derivatives, and are thus Lagrange multipliers for the constraints
\begin{equation}
C^m \equiv  -2 \partial_{n}\pi^{mn}=0 \ \ \ \ \ C \equiv  -\Delta h+\partial_{m}\partial_{n}h^{mn}=0.
\end{equation}

The constraints generate the gauge symmetries (linearized diffeomorphisms) through the Poisson brackets.  These are, in terms of the canonical variables,
\begin{equation}
\pi^{mn}\rightarrow \pi^{mn}-\partial^{m}\partial^{n}\xi^{0}+\delta^{mn}\xi^{0} \label{b}
\end{equation}
\begin{equation}
h_{mn}\rightarrow h_{mn}+\partial_{m}\xi_{n}+\partial_{n}\xi_{m}
\end{equation}

\subsection{Solving the momentum constraint $C_m= 0$}

By introducing the double dual $\Pi_{ijk pqr} =  \epsilon_{ijkm} \epsilon_{pqrn} \pi^{mn}$ of the momentum $\pi^{mn}$, which is a tensor of $(3,3)$ Young symmetry type, the constraint $C_m = 0$ is seen to be equivalent to 
$$
\partial_{[l}\Pi_{ijk] pqr}  = 0,
$$
or, in the language of \cite{DVH}
$$
d \Pi = 0.
$$
Here $d$ is a differential operator for $2$-columns tensors that has the property $d^3 = 0$. The Poincar\'e lemma for $d$ demonstrated in  \cite{DVH} implies then
$$
\Pi = d^2 P
$$
for some $P$ of 
$$(2,2) \equiv \yng(2,2) $$
Young symmetry type.  The tensor $P$ itself is determined from $\Pi$ up to the $d$ of a $(2,1)$-tensor \cite{DVH},
$$
\delta_1 P = d \chi
$$
where $\chi$ is of $(2,1)$ Young symmetry type.  We call $P$ the prepotential for the conjugate momentum.

In components, these formulas read (absorbing numerical factors in redefinitions)
\begin{equation}
\pi^{mn}=\epsilon^{mkqs}\epsilon^{nrtu} \partial_{k}\partial_{r}P_{qstu} \label{piPe}
\end{equation}
and
$$
\delta_{1}P_{qstu}=\chi_{tu[q,s]}+\chi_{qs[t,u]}
$$
with $\chi_{stu} = - \chi_{tsu}$, $\chi_{[stu]} = 0$.

Furthermore, the transformation on the prepotential that induces the gauge transformation (\ref{b}) on the momentum is easily verified to be the ``Weyl-type" transformation :
\begin{equation}
\delta_{c}P^{qstu}=\frac{1}{4}\left[\delta_{qt}\delta_{su}-\delta_{qu}\delta_{st}\right]\xi.
\end{equation} 
The total gauge transformation on $P$ is then:
\begin{equation}
\delta P_{qstu}=\delta_{1}P_{qstu}+\delta_{c}P_{qstu}\label{gauge1}
\end{equation}

A straightforward computation shows that the momentum $\pi^{mn}$ conjugate to the metric is equal to $-G^{mn}[P]$,
\be
\pi^{mn} = - G^{mn}[P],
\ee
where $G^{mn}[P]$ is the Einstein tensor of $P_{ijrs}$ defined in appendix \ref{Cotton22}. So,  the resolution of the momentum constraint amounts to equate $\pi^{mn}$ to the Einstein tensor of a $(2,2)$-prepotential, which identically fulfills $\partial_m G^{mn}=0$ by the contracted Bianchi identity.

\subsection{Solving the Hamiltonian constraint $C=0$}

Following \cite{Henneaux:2004jw}, we  decompose the spatial components of the Pauli-Fierz field as follows:

\begin{equation}
h_{mn}=j_{mn}+\partial_{m}u_{n}+\partial_{n}u_{m}\label{h}
\end{equation}
with $j_{mn}$ traceless. Substituting this expression in the constraint, we get $\partial_{m}\partial_{n}j^{mn}=0$.  Double-dualization of this equation yields $d^2 \Sigma = 0$ for the double-dual
$$ \Sigma_{ijkpqr} = \epsilon_{ijkm} \epsilon_{pqrn} j^{mn}$$ of Young symmetry 
$$(3,3) \equiv \yng(2,2,2).$$  Here, $d$ is again the operator of \cite{DVH} which fulfills $d^3 = 0$. The results of \cite{DVH} imply then $\Sigma = d M$ where $M$ is a tensor of Young type $(3,2)$.  In components,
\begin{equation}
j_{mn}=\epsilon_{nkst}\partial^{k}\Phi_{\ \   m}^{ st}+\epsilon_{mkst}\partial^{k}\Phi_{\ \ n}^{ st} \label{jPhi}
\end{equation}
where 
$$
\Phi^{\ \ m}_{st} = \frac{1}{3!}\epsilon^{mijk} M_{ijkst}
$$
The tensor $\Phi^{\ \ m}_{ st}$ is traceless, $\Phi^{\ \ \ m}_{sm}= 0$, since $M_{[ijks]t} = 0$. However, if non-zero, the trace of $\Phi^{\ \ m}_{st}$ would in any case drop from (\ref{jPhi}),  so we shall allow it to be non-zero, accounting for this possibility by introducing the gauge invariance
$$
\delta_{C}\Phi_{rsm}=B_{\left[\right. r}\delta_{\left. s\right] m}
$$

There is one more condition on $\Phi_{rsm}$ which follows from the tracelessess of $j^{mn}$.  It is $\partial_{[k} \Phi_{ijm]} = 0$, from which one infers that the totally antisymmetric part $ \Phi_{[ijm]}$ of $ \Phi_{ijm}$ is of the form $\partial_{[i} \lambda_{jm]}$ and can thus be absorbed in a redefinition of the vector $u_m$ in (\ref{h}).   We can thus assume $ \Phi_{[ijm]}= 0$, so  that the prepotential $ \Phi_{ijm}$ for $h_{mn}$ is of Young symmetry type $(2,1)$.

In order to obtain the associated gauge symmetries, we follow the same procedure as in \cite{Henneaux:2004jw} for the four dimensional case. We  arrive in this way at
\begin{equation}
\delta\Phi_{mrs}=\delta_{C}\Phi_{mrs}+\delta_{1}\Phi_{mrs}\label{gauge2}
\end{equation}
with 
\begin{equation}
\delta_{1}\Phi_{mrs}=\partial_{r}S_{sm}-\partial_{s}S_{rm}+\partial_{r}A_{sm}-\partial_{s}A_{rm}+2\partial_{m}A_{sr}
\end{equation}
where $S_{sr}$ is symmetric, while $A_{sr}$ is antisymmetric. 

\subsection{Action in terms of prepotentials}
\subsubsection{Kinetic term}
With the above expressions, one finds that the Hamiltonian kinetic term $\int dx^0 d^4x \pi^{mn} \dot{h}_{mn}$ is equal to 
\[
\int dtd^{4}x \pi^{mn}\dot{h}_{mn}=\int dtd^{4}x\partial_{p}\partial_{r}\epsilon^{mpqs}\epsilon^{nrtu}P_{qstu}\partial_{k}[\epsilon_n^{\; \; kab}\dot{\Phi}_{abm}+\epsilon_m^{\; \; kab}\dot{\Phi}_{ abn}]=
\]
\[
=2\int dt d^{4}x\partial_{p}\partial_{r}\epsilon^{mpqs}\epsilon^{nrtu}P_{qstu}\partial_{k}\epsilon_n^{\; \; kab}\dot{\Phi}_{ abm}
\]
Using the formulas in Appendix  \ref{WeylT},  this term can also be rewritten as
\be 
\int dx^0 d^4x \pi^{mn} \dot{h}_{mn} =-4 \int dx^0 d^4x D_{ijm}[P] \dot{\Phi}^{ijm}  \label{HKTpf1}
\ee
where $D_{ijm}[P]$ is the co-Cotton tensor of $P_{ijrs}$. This form of the kinetic term makes it obvious that it is invariant under the gauge symmetries of the prepotential $P_{ijrs}$ (since the co-Cotton tensor is), as well as the gauge symmetries of the prepotential  $\Phi_{ijm}$.  This is because  the co-Cotton tensor is traceless and divergence-free.

Equivalently, one can rewrite the kinetic term as 
\be 
\int dx^0 d^4x \pi^{ij} \dot{h}_{ij} =-4 \int dx^0 d^4x D_{ijmn}[\Phi] \dot{P}^{ijmn} \label{HKTpf2}
\ee
where $D_{ijmn}[\phi]$ is the co-Cotton tensor of $\Phi_{ijm}$, which is also divergence-free and, in this case, doubly-traceless.

\subsubsection{Hamiltonian}

Expressing the Pauli-Fierz field and its conjugate momentum in terms of the prepotentials yield the Hamiltonian
\be
H = \int d^3x \left( R^{ij}[P]R_{ij} [P] - \frac{7}{27} R^2[P] + 2 E^{ijk}[\Phi] E_{ijk}[\Phi] - \frac{3}{2} E^i[\Phi] E_i[\Phi] \right) \label{HamPrep000}
\ee
 The Hamiltonian is clearly invariant under the ``diffeomorphisms" of the prepotentials since it involves only their curvatures.  It is also invariant under the ``Weyl rescalings" up to a divergence, as one can easily verify.


\section{Prepotentials - Starting from the Curtright action}
\setcounter{equation}{0}
\label{PrepCurt}

One finds the same kinetic term and Hamiltonian if one starts from the Curtright action, as we now show.

\subsection{Hamiltonian formulation}
The momenta  conjugate to the spatial components of the Curtright field are easily found to be
\begin{equation}
\pi^{ijk}=\frac{\partial{\cal{L}}}{\partial\dot{T}_{ijk}}= V^{ijk} -\delta^{jk}\, V^{il}_{\ \ l}  +\delta^{ik}V^{jl}_{\ \ l}\label{mom}
\end{equation}
in terms of the invariant velocities.  We can solve for the velocities to get
$$
V^{ijk}=\pi^{ijk}-\frac12 \delta^{jk}\,\pi^{il}_{\ \ l} + \frac12 \delta^{ik}\, \pi^{jl}_{\ \ l},
$$
yielding the action in canonical form,
\begin{equation}
S[T_{ijk}, \pi^{ijk}, m_i, m_{jk}]=\int dt \ d^{4}x \ \left[\pi_{ijk}\dot{T}^{ijk}-{\cal{H}} - m_j \, \Gamma^j - m_{jk} \Gamma^{jk} \right]
\end{equation}
with $m_j \equiv 2 T_{0j0}$  and $m_{jk} \equiv T_{0jk}$.  Here the Hamiltonian ${\cal{H}}$ reads
\[
{\cal{H}}=\frac{1}{2}\pi_{ijk}\pi^{ijk}-\frac{1}{2}\pi_{ik}^{\ \ k}\pi^{il}_{\ \ l}+\frac{1}{2}\partial_{i}T_{jkl}\partial^{i}T^{jkl}+\partial_{i}T_{jkl}\partial^{j}T^{kil}-\partial_{i}T_{jk}^{\ \ k}\partial^{i}T^{jl}_{\ \ l}-
\]
\[
-\partial_{i}T_{jk}^{\ \ k}\partial^{j}T^{li}_{\ \ l}-2\partial_{i}T_{jl}^{\ \ l}\partial^{k}T^{ij}_{\ \ k}-\frac{1}{2}\partial_{l}T_{ij}^{\ \ l}\partial^{k}T^{ij}_{\ \ k}
\]
while $\Gamma^{jk}$ and $\Gamma^j$ are respectively given by 
\begin{equation}
\Gamma^{jk} \equiv -2 \partial^{i}(\pi_{ijk}+\pi_{kji})
\end{equation}
and
\begin{equation}
\Gamma^j \equiv \Delta T^{jk}_{\ \ k}+\partial_{i}\partial^{j}T^{ki}_{\ \ k}+\partial_{i}\partial^{k}T^{ij}_{\ \ k}
\end{equation}

The temporal $T_{j00}$ and $T_{i0j}$ may thus be regarded as Lagrange multipliers imposing the constraints:
\begin{equation}
\Gamma^{jk} =0 \label{cons}
\end{equation}
and
\begin{equation}
\Gamma^j  =0 \label{consBis}
\end{equation}

The brackets between the Hamiltonian variables are
\begin{eqnarray}
&& \{T_{ijk} (\vec{x}), \pi^{mnp} (\vec{y})\} = \nonumber \\
&& \hspace{.4cm} \frac13\left( \delta^m_i \delta ^n_j \delta^p_k - \delta^n_i \delta ^m_j \delta^p_k  + \frac12 (\delta^m_i \delta ^p_j \delta^n_k - \delta^p_i \delta ^m_j \delta^n_k + \delta^p_i \delta ^n_j \delta^m_k - \delta^n_i \delta ^p_j \delta^m_k) \right) \hspace{.6cm} \label{brackets}
\end{eqnarray}
and such that 
$$  \{T_{ijk} (\vec{x}), \int d^4 x\pi^{mnp} (\vec{y})u_{mnp}(\vec{y})\} = u_{ijk}(\vec{x})^P,
$$
where $ u_{ijk}(\vec{x})^P$ is the standard  projection of the tensor $ u_{ijk}(\vec{x})$ on the subspace with Young symmetry $(2,1)$. From (\ref{brackets}) follows
$$\{T_{ik}^{\ \ k} (\vec{x}), \pi^{mnp} (\vec{y})\} = \frac12 \left(\delta_i^m \delta^{np} - \delta_i^n \delta^{mp} \right).
$$
 The dynamical  Hamiltonian equations take the form
$$
 \dot{F} = \{F, \int d^4 x \left( {\cal{H}} + m_j \, \Gamma^j + m_{jk} \Gamma^{jk} \right) \}.
$$

The Hamiltonian constraints are in fact the canonical generators of the gauge symmetries of the theory.  From (\ref{GaugeInvVel}) one finds
$$
\delta\pi^{ijk}=\partial^{i}\partial^{k}(\si^{0j}-3\al^{0j})-\partial^{j}\partial^{k}(\si^{0i}-3\al^{0i})-
$$
\[
-\delta^{jk}[\partial^{i}\partial_{l}(\si^{0l}-3\al^{0l})-\Delta(\si^{0i}-3\al^{0i})]+\delta^{ik}[\partial^{j}\partial_{l}(\si^{0l}-3\al^{0l})-\Delta(\si^{0j}-3\al^{0j})],
\] a transformation that coincides with the transformation generated by $\int d^4 x \xi_j \Gamma^j$,
\begin{eqnarray}
&&\{\delta\pi^{ijk}, \int d^4 y \,  \xi_m \, \Gamma^m\} =  \frac12\left( \partial^{j}\partial^{k}\xi^{i}- \partial^{i}\partial^{k}\xi^{j} \right) \nonumber \\ 
&& \hspace{1cm} + \frac12 \left( \delta^{jk}(\partial^i \partial_m \xi^m - \triangle \xi^i) -  \delta^{ik}(\partial^j \partial_m \xi^m - \triangle \xi^j)\right) \label{Gaugepi2}
\end{eqnarray}
 if one identifies the gauge parameter $\xi_i$ with  $2(\si_{0i}-3\al_{0i})$.  Similarly, the linear combination $\xi_{jk} \Gamma^{jk}$ generates 
\begin{eqnarray} \delta T_{ijk} &=& \{T_{ijk}, \int d^4x \, \xi_{mn} \Gamma^{mn} \} \nonumber\\
&=& \partial_i \si_{jk}- \partial_j \si_{ik} + \partial_i \al_{jk}- \partial_j \al_{ik} - 2  \partial_k \al_{ij}
\end{eqnarray}
with $\si_{mn} = \xi_{(mn)}$, $\al_{mn} = \xi_{[mn]}$.

If we recalls that the free (2,1) theory is the dual theory for linearized gravity, it seems reasonable to conjecture at this point that these constraints may be solved in terms of two prepotentials featuring the mixed symmetry (2,1) and (2,2), respectively.  This indeed turns out to be the case as we now show.

\subsection{Solving the momentum constraints $\Gamma_{mn}=0$}
The momentum constraints are easily seen to be equivalent to 
$$
\partial_i \pi^{ijk} = 0.
$$
and imply
$$
\partial_i \pi^{jki} = 0.
$$
Introducing the double-dual $A_{rst mn}$ of $\pi^{ijk}$, which is a tensor of $(3,2)$-type
$$
\pi^{ijk} = \frac12 \frac{1}{3!} \epsilon^{ijmn} \epsilon^{krst} A_{rstmn} \; \; \Leftrightarrow \; A_{rstmn} = - \frac12 \epsilon_{rstk} \epsilon_{mnij} \pi^{ijk},
$$
one may rewrite these equations as
$$
\partial_{[p} A_{rst]mn} = 0 , \; \; \; \; \; \; A_{rst [mn,p]} = 0.
$$
This implies \cite{Coho}
$$
A_{rstmn} = - 6 \partial_{[r}\Psi_{st][m,n]}
$$ 
for some ``prepotential" $\Psi_{rsm}$ which has the $(2,1)$ Young symmetry. The factor $(-6)$ has been inserted to avoid numerical factors in (\ref{piPhi}) below. In terms of the conjugate momenta, one gets
\begin{equation}
\pi^{ijk} =  \epsilon^{ijmn} \epsilon^{krst} \partial_r \partial_m \Psi_{stn}  \label{piPhi0}
\ee 
The prepotential $\Psi_{rsm}$ is determined from $\pi^{ijk}$ up to 
\begin{equation}
\delta_{1}\Psi_{mrs}=\partial_{r}s_{sm}-\partial_{s}s_{rm}+\partial_{r}a_{sm}-\partial_{s}a_{rm}+2\partial_{m}a_{sr}
\end{equation}
where $s_{sr}$ is symmetric, while $a_{sr}$ is antisymmetric. Furthermore, the transformation of the prepotential that accounts for the gauge transformations (\ref{Gaugepi2}) of the momenta  are again the Weyl rescalings
\begin{equation}
\delta_{C}\Psi_{rsm}=\xi_{\left[\right. r}\delta_{\left. s\right] m}
\end{equation}

We thus see that the momentum constraint of the dual formulation can be solved in terms of a prepotential $\Psi_{rsm}$ with the same  Young symmetry $(2,1)$ as the prepotential $\Phi_{rsm}$ needed to solve the Hamiltonian constraint of the original  Pauli-Fierz formulation.  In addition, both prepotentials have the same gauge symmetries.  This enables one to identify them,
\be
\Psi_{rsm} = \Phi_{rsm}. 
\ee

Note that one may identify $\pi^{ijk}$ with the Einstein tensor of the prepotential (up to the numerical factor $-2$),
\be
\pi^{ijk} =-2  G^{ijk}[\Phi], \label{piPhi}
\ee
in a manner similar to what we found for the Pauli-Fierz conjugate momentum.

\subsection{Solving the Hamiltonian constraints $\Gamma^{j}=0$}

To solve the Hamiltonian constraints  $\Gamma^{j}\equiv \Delta T^{jk}_{\ \ k}+\partial_{i}\partial^{j}T^{ki}_{\ \ k}+\partial_{i}\partial^{k}T^{ij}_{\ \ k}=0$ of the Curtright field, we proceed in a manner that parallels the procedure followed in \cite{Henneaux:2004jw} for solving the Hamiltonian constraint of the Pauli-Fierz field.

The starting point is to notice that it is always possible to decompose the field $T_{ijk}$ as
\be
T_{ijk}=t_{ijk}+ \theta_{ijk} \label{DecompT}
\ee
where $t_{ijk}$ is traceless, $t_{ik}^{\ \ k}=0$, and where $\theta_{ijk}$ is a pure gauge term carrying the trace,
\[
\theta_{ijk}= \partial_i u_{jk}- \partial_j u_{ik} + \partial_i v_{jk}- \partial_j v_{ik} - 2  \partial_k v_{ij}
\]
with $u_{mn} = u_{(mn)}$, $v_{mn} = v_{[mn]}$.  The fields $u_{ij}$ and $v_{ij}$ are prepotentials for the Curtright fields analogous to the prepotential $u_i$ for the Pauli-Fierz field $h_{ij}$ and drop out from the subsequent formulas on account of gauge invariance.  The relevant piece of the Curtright field is $t_{ijk}$, which we shall now express in terms of a prepotential $P_{ijkl}$ possessing the algebraic symmetries of a $(2,2)$ Young tensor.

The Hamiltonian constraint $\Gamma^{j}=0$ becomes the equation $\partial_i \partial_k t^{ijk} = 0$ for $t^{ijk}$.  It is shown in appendix \ref{SHCCT} that the general solution to  this equation can be assumed to take the form
\begin{equation}
t^{ijk}=-\frac{2}{3}\partial_{l}\left[2\epsilon^{klab}P_{[ab]}^{\ \ \ [ij]}+\epsilon^{ilab}P_{[ab]}^{\ \ \ [kj]}-\epsilon^{jlab}P_{[ab]}^{\ \ \ [ki]}\right] \label{tP}
\end{equation}
where the prepotential $P_{[ab][cd]}$  transforms in the irreducible representation $(2,2)$,
$$ P_{abcd} = P_{[ab][cd]}, \; \; \; P_{[abc]d} = 0.$$

One can easily check that the following transformations  for the prepotential $P_{abcd}$ leave $T_{ijk}$ invariant up to a gauge transformation:
\be 
\delta P_{abcd} = \delta_{1}P_{abcd} + \delta_{C}P_{abcd} \label{gaugeP222}
\ee
with
\begin{equation}
\delta_{1}P_{abcd}=\chi_{cd[a,b]}+\chi_{ab[c,d]},
\end{equation}
where $\chi_{abc} = - \chi_{bac}$ transforms in the $(2,1)$-representation, i.e., $\chi_{[abc]} = 0$, and
\begin{equation}
\delta_{c}P_{abcd}=\frac{1}{4}[\delta_{ac}\delta_{bd}-\delta_{ad}\delta_{bc}]\xi.
\end{equation}
Conversely, given the tensor $T_{ijk}$ up to a gauge transformation, i.e., its curvature $E_{ijkmn}$ -- or what is the same in four dimensions, its Ricci tensor $E_{ijk}$ --, one knows the Cotton tensor of  the $(2,2)$-prepotential $P_{ijkl}$ (see section \ref{Cotton22}) since the Ricci tensor of $T_{ijk}$ is easily verified to be equal to the co-Cotton tensor of $ P_{ijkl}$ (see formula (\ref{TSchoutenP}) below).
Hence $P_{ijkl}$ is determined up to the transformations (\ref{gaugeP222}), which are therefore  its ``gauge symmetries".

We thus see that the Curtright field $T_{ijk}$  can be expressed in terms of the same prepotential, with same gauge symmetries, as the conjugate momentum $\pi^{ij}$ of the Pauli-Fierz field.   This enables one to identify the prepotential for $T_{ijk}$ with the prepotential for $\pi^{ij}$.

\subsection{Action in terms of prepotentials}

\subsubsection{Kinetic term}

The Hamiltonian kinetic term of the Curtright theory is (up to total derivatives)
\[
\int dt d^{4}x \pi_{ijk}\dot{T}^{ijk}=\int dt d^{4}\pi_{ijk}\dot{t}^{ijk}=-2\int dt d^{4}x \partial^{l}\partial^{m}\epsilon_{ijln}\epsilon_{kmab}\Phi^{abn}\partial_{p}\epsilon^{kpcd}\dot{P}^{ij}_{cd}=
\]
\[
=2\int dt d^{4}x \partial^{l}\partial^{p}\epsilon_{nlij}\epsilon_{mpcd}P^{ij}_{cd}\partial^{k}\epsilon_{mkab}\dot{\Phi}^{abn}.
\]
This expression has exactly the same form as the Hamiltonian kinetic term emerging from the Pauli-Fierz Lagrangian.  Hence, one can also write it as in (\ref{HKTpf1}) and (\ref{HKTpf2}),
\begin{eqnarray} 
\int dt d^{4}x \pi_{ijk}\dot{T}^{ijk} &=&-4 \int dx^0 d^4x D_{ijm}[P] \dot{\Phi}^{ijm} \\
&=& -4 \int dx^0 d^4x D_{ijmn}[\Phi] \dot{P}^{ijmn}.
\end{eqnarray}

\subsubsection{Hamiltonian}

The kinetic energy density 
$\frac{1}{2}\pi_{ijk}\pi^{ijk}-\frac{1}{2}\pi_{ij}^{\ \ j}\pi^{il}_{\ \ l}$ of the Curtright field becomes, when expressed in terms of the prepotential $\Phi_{ijk}$,
\[
\frac{1}{2}\pi_{ijk}\pi^{ijk}[\Phi]-\frac{1}{2}\pi_{ij}^{\ \ j}\pi^{il}_{\ \ l}[\Phi]=
\]
\[
=2\Delta\Phi_{prs}\Delta\Phi^{prs}-4\partial_{u}\partial^{i}\Phi_{pil}\partial^{u}\partial_{k}\Phi^{pkl}-2\Delta\Phi^{n}_{\ ns}\Delta\Phi_{m}^{\ ms}-2\partial_{u}\partial_{i}\Phi^{i}_{\ rs}\partial^{u}\partial^{m}\Phi_{m}^{\ rs}+
\]
\[
+2\partial_{u}\partial^{k}\Phi^{n}_{\ nk}\partial^{u}\partial_{i}\Phi_{m}^{\ mi}+4\Delta\Phi_{n}^{\ ns}\partial^{i}\partial^{m}\Phi_{mis}+2\partial_{n}\partial_{i}\Phi^{nil}\partial^{p}\partial^{k}\Phi_{pkl}
\]
But this is precisely equal to the potential energy density $\frac{1}{4}\partial_{i}h_{jk}\partial^{i}h^{jk}[\phi]-\frac{1}{2}\partial_{i}h_{jk}\partial^{j}h^{ik}[\phi]$  of the Pauli-Fierz field (up to total derivative), and so one has
\begin{eqnarray} 
&& \hbox{\bf Kinetic Energy of Curtright field} = \nonumber \\
&& \hspace{3.5cm} \hbox{\bf Potential Energy of Pauli-Fierz field} \hspace{2cm}\\
&& \hspace{3.5cm} =  \int d^3x \left( 2 E^{ijk}[\Phi] E_{ijk}[\Phi] - \frac{3}{2} E^i[\Phi] E_i[\Phi] \right)
\end{eqnarray}
(using (\ref{HamPrep000})).

Similarly, one finds,
\[
\frac{1}{2}\partial_{i}t_{jkl}\partial^{i}t^{jkl}[P]+\partial_{i}t_{jkl}\partial^{j}t^{kil}[P]-\frac{1}{2}\partial^{k}t_{ijk}\partial_{l}t^{ijl}[P] =
\]
\[
\pi_{mn}\pi^{mn}[P]-\frac{\pi^{2}}{3}[P]
\]
(up to total derivatives) and so
\begin{eqnarray} 
&& \hbox{\bf Potential Energy of Curtright field} = \nonumber \\
&& \hspace{3.5cm} \hbox{\bf Kinetic Energy of Pauli-Fierz field} \hspace{2cm}\\
&& \hspace{3.5cm} =  \int d^3x \left(  R^{ij}[P]R_{ij} [P] - \frac{7}{27} R^2[P]\right)
\end{eqnarray}
(using (\ref{HamPrep000})).

This completes the proof that both theories have identical actions when they are written in terms of the prepotentials.  Note that the interchange  of the kinetic and potential energies occurs already for $p$-forms gauge fields, in particular in the Maxwel theory, where the electric field squared is the kinetic energy in the electric representation, and the potential energy in the magnetic representation (and vice-versa for the magnetic field).


\section{Symplectic Structure}
\setcounter{equation}{0}
\label{PBStructure}

The prepotentials $\Phi_{ijk}$ and $P_{ijkl}$ for the  fields $h_{ij}$ and $T_{ijk}$ are canonically conjugate in the sense that while $\Phi_{ijk}$ is the prepotential for the Pauli-Fierz field $h_{ij}$, the prepotential $P_{ijkl}$ is  the prepotential for its conjugate momentum $\pi^{ij}$.  

It is interesting to exhibit the canonical structure in terms of the dual fields $h_{ij}$ and $T_{ijk}$ rather than the prepotentials. It turns out to be actually more convenient to compute the brackets  between the gauge invariant quantities $R_{ijkl}[h](\vec{x})$ and $E_{mnpqr}[T](\vec{y})$, as we now discuss.

Before doing this, we derive a number of useful relations between the variables appearing in the formalism.

\subsection{The ``graviton set"}
As we have just established, the graviton field $h_{ij}$ can be equivalently described in terms of the prepotential $\Phi_{ijk}$ or the momentum conjugate to the Curtright field,
\be h_{ij} \; \;  \leftrightarrow \; \;  \Phi_{ijk}   \; \;  \leftrightarrow \; \;  \pi^{ijk}. \label{GravSet} \ee 
The graviton field $h_{ij}$ is subject to the Hamiltonian constraint ${\mathcal H} = 0$, while the momentum conjugate to the Curtright field  is subject to the momentum constraints $\Gamma_{ij} = 0$.  The prepotential $\Phi_{ijk}$ is subject to no constraint.  

We shall call the variables appearing in (\ref{GravSet}), the ``variables of the graviton set".  We have derived explicit expressions for $h_{ij}$ and $\pi^{ijk}$ in terms of the prepotential $ \Phi_{ijk}$ (formulas (\ref{h})-(\ref{jPhi}) and (\ref{piPhi}), respectively).  

From these relations,  one easily derives the useful identities
\be
\pi_{ijk} - \frac12 \delta_{jk} \pi_{il}^{\; \; l} + \frac12 \delta_{ik} \pi_{jl}^{\; \; l} = -2 S_{ijk} [\Phi]
\ee
and
\be
{\mathcal E}_{rsij}[h]=  2 D_{rsij}[\Phi].  
\ee
Here, $S_{ijk}[\Phi]$ and $D_{rsij}[\Phi]$ are respectively the Schouten and co-Cotton tensors for the tensor $\Phi_{ijm}$ of $(2,1)$ Young symmetry type introduced in appendix \ref{Weyl01}.

The relations (\ref{h})-(\ref{jPhi}) and (\ref{piPhi}) can be inverted to yield the prepotential $ \Phi_{ijk}$ in terms of the fields $h_{ij}$ or $\pi^{ijk}$ (obeying their constraints).  The explicit inversion formulas are non local and involves gauge choices. They are given in the appendix \ref{AppD}.  From these formulas, one can in principle express the Pauli-Fierz field $h_{ij}$ in terms of the momentum $\pi^{ijk}$ conjugate to the Curtright field and vice-versa.  This will not be done explicitly here as it is not needed.

\subsection{The conjugate ``dual graviton set"}
Similarly, the Curtright field $T_{ijk}$ of the dual description can be equivalently described in terms of the conjugate prepotential $P_{ijkl}$ or the momentum $\pi^{ij}$ conjugate to the graviton field, 
\be T_{ijk} \; \;  \leftrightarrow \; \;  P_{ijkl}   \; \;  \leftrightarrow \; \;  \pi^{ij}. \label{DualGravSet} \ee The dual graviton field $T_{ijk}$ is subject to the Hamiltonian constraints $\Gamma_i = 0$, while the  conjugate momentum $\pi^{ij}$  is subject to the momentum constraints ${\mathcal H}_i = 0$.  The prepotential $P_{ijkl}$ is subject to no constraint.  We shall call the variables appearingg in (\ref{DualGravSet}), the ``variables of the dual graviton set".

Useful identities that follow from the relations (\ref{tP}) and (\ref{piPe}) are:
\be
- \pi_{in} + \frac{\pi}{3}\delta_{in} = S_{in}[P]  \label{piSchoutenP}
\ee
and
\be
{\mathcal E}_{rbi}[T]= -2 D_{rbi}[P]   \label{TSchoutenP} 
\ee
where $S_{in}[P]$ and $D_{rbi}[P]$ are respectively the Schouten and co-Cotton tensors of $P_{ijmn}$ introduced in appendix \ref{Cotton22}.  

The relations (\ref{tP}) and (\ref{piPe}) can also be inverted to yield the prepotential $ P_{ijmn}$ in terms of the fields $T_{ink}$ or $\pi^{ij}$ (obeying their constraints).  The explicit inversion formulas are again non local and involves gauge choices. They are also given in the appendix \ref{AppD}.  From these formulas, one can express the Curtright field $T_{ijk}$ in terms of the momentum $\pi^{ij}$ conjugate to the Pauli-Fierz field and vice-versa.  One can then derive the expression of  $E_{abcde}[T]$ in terms of $\pi^{ij}$.  One gets explicitly
\begin{eqnarray}
3E_{abcde}[T]&=&- 2\left(\epsilon_{abcm}\partial_{d}\pi^{m}_{\ e}-\epsilon_{abcm}\partial_{e}\pi^{m}_{\ d} \right)
\nonumber \\
&& \hspace{.1cm}
-\partial_{d}\epsilon_{caem}\pi^{m}_{\ b} +\partial_{e}\epsilon_{cadm}\pi^{m}_{\ b}-\epsilon_{abem}\partial_{d}\pi^{m}_{\ c} \nonumber \\
&& + \epsilon_{abdm}\partial_{e}\pi^{m}_{\ c}-\epsilon_{bcem}\partial_{d}\pi^{m}_{\ a}+\epsilon_{bcdm}\partial_{e}\pi^{m}_{\ a}
\label{EEpi}
\end{eqnarray}
The right-hand side is easily seen to be such that $E_{abcde}$ obeys the identities $\partial_{[m}E_{abc]de}= 0$ and $E_{abc[de,m]} = 0$ because $\pi^{ij}$ is transverse, $\partial_i \pi^{ij} = 0$.

\subsection{Poisson bracket of the curvatures in $D=4$}

We can now turn to the commutation of the Poisson bracket structure. To warm up, we first compute the Poisson Bracket $\{G_{ab}[h],G_{cd}[f]\}$ of the Einstein tensor of the dual metric s $h_{ij}$ and $f_{ij}$ of the duality-invariant formulation of linearized gravity in four space-time dimensions, where the dual field to $h_{ij}$ is also a symmetric tensor $f_{ij}$. Explicitly,
\[
G_{ab}[h]=\frac{1}{2}\left[\partial_{a}\partial^{m}h_{mb}+\partial_{b}\partial^{m}h_{ma}-\Delta h_{ab}-\partial_{a}\partial_{b}h\right]-\frac{1}{2}\left[\partial^{m}\partial^{n}h_{mn}-\Delta h\right]\delta_{ab}
\]
\[
G_{cd}[f]=\frac{1}{2}\left[\partial_{c}\partial^{m}f_{md}+\partial_{d}\partial^{m}f_{mc}-\Delta f_{cd}-\partial_{c}\partial_{d}f\right]-\frac{1}{2}\left[\partial^{m}\partial^{n}f_{mn}-\Delta f\right]\delta_{cd}
\]
We recall that in three dimensions, the Riemman tensor is completely captured by the Einstein tensor.

Our strategy is to start from the canonical bracket 
\[
\{h_{ij}(\vec{x}),\pi_{mn}(\vec{y})\}=\frac{1}{2}\left[\delta_{im}\delta_{jn}+\delta_{in}\delta_{jm}\right]\delta^{(3)}(\vec{x}-\vec{y})
\]
and express $f_{ij}$ in terms of the canonical momentum $\pi^{ij}$ conjugate to $h_{ij}$. From\begin{equation}
f_{ij}=\partial^{r}\epsilon_{irs}P^{s}_{j}+\partial^{r}\epsilon_{jrs}P^{s}_{i}+\partial_{i}v_{j}+\partial_{j}v_{i} \label{f}
\end{equation}
one finds $G[f(P)]$:
\begin{equation}
G_{cd}[f(P)]=\frac{1}{2}\left[\partial_{c}\partial^{m}\epsilon_{drs}\partial^{r}P^{s}_{m}+\partial_{d}\partial^{m}\epsilon_{crs}\partial^{r}P^{s}_{m}-\Delta\epsilon_{crs}\partial^{r}P^{s}_{d}-\Delta\epsilon_{drs}\partial^{r}P^{s}_{c}\right]
\end{equation}
yielding, since
\begin{eqnarray}
\pi^{ij} &=&\epsilon^{ipq}\epsilon^{jrs}\partial_{p}\partial_{r}P_{qs}  \nonumber \\
&=& \delta^{ij}(\Delta P-\partial^{m}\partial^{n} P_{mn})-\partial^{i}\partial^{j}P+\partial^{i}\partial_{r}P^{jr}+\partial^{j}\partial_{r}P^{jr}-\Delta P^{ij}\label{mt}, \hspace{1cm}
\end{eqnarray}
the relation
\begin{equation}
G_{cd}[f(\pi)]=\frac{1}{2}\left[\partial^{r}\epsilon_{drs}\pi_{c}^{s}+\partial^{r}\epsilon_{crs}\pi_{d}^{s}\right]
\end{equation}

Now it is straightforward to compute the Poisson bracket $\{G_{ab}[h](\vec{x}),G_{cd}[f](\vec{y})\}$.  One obtains
\begin{equation}
\{G_{ab}[h(x)],G_{cd}[f(y)]\}=\int d^{3}z\frac{\delta G_{ab}[h]}{\delta h^{pq}(z)}\frac{\delta G_{cd}[f]}{\delta\pi_{pq}(z)}=
\end{equation}
\[
=\frac{1}{8}\left[\partial_{a}\partial_{c}\partial^{r}\epsilon_{drb}+\partial_{a}\partial_{d}\partial^{r}\epsilon_{crb}+\partial_{b}\partial_{c}\partial^{r}\epsilon_{dra}+\partial_{b}\partial_{d}\partial^{r}\epsilon_{cra}-\right.
\]
\[
\left.-\Delta\partial^{r}\epsilon_{drb}\delta_{ca}-\Delta\partial^{r}\epsilon_{dra}\delta_{cb}-\Delta\partial^{r}\epsilon_{crb}\delta_{da}-\Delta\partial^{r}\epsilon_{cra}\delta_{bd}\right]\delta^{(3)}(\vec{x}-\vec{y})
\]
where all derivatives are taken with respect to $x$. It should be noted that the right-hand side of this relation is identically divergence-free (with respect to $a$, $b$, $c$ or $d$), as it should since the left-hand side is.

\subsection{Poisson brackets of the curvatures in $D=5$}

The computation can be extended to cover the 5-dimensional case by following the same lines.  Equation (\ref{EEpi})  gives the curvature $E_{abcde}[T]$ in terms of the momentum $\pi^{ij}$ conjugate to the metric $h_{ij}$.  From this expression, it is direct, although somewhat tedious, to derive the bracket $\{R_{ijkl}[h](\vec{x}), E_{mnprs}[T](\vec{y})\} $.  One finds
\begin{eqnarray}
&&\{R_{ijmn}[h](\vec{x}),E_{abcde}[T](\vec{y})\} = -\frac{2}{3}\Big[2 \left(\epsilon_{abc[n}\partial_{m]}\partial_{[j}\delta_{i][e}\partial_{d]} 
+\epsilon_{abc[j}\partial_{i]}\partial_{[n}\delta_{m][e}\partial_{d]}\right)  \nonumber \\
&& +  \left( \partial_{[e}\epsilon_{d]ab[i} \partial_{j]}  \delta_{c[n} \partial_{m]} 
+ \partial_{[e}\epsilon_{d]ab[m} \partial_{n]}  \delta_{c[j} \partial_{i]}
 + \sum \hbox{cyclic} \right)\Big] \delta^{3}(\vec{x}-\vec{y})
\end{eqnarray}
where antisymmetrization is always over a pair of adjacent indices and where the sum is over the 3 cyclic permutations of $(a,b,c)$.
The right-hand side of this expression is easily verified  to fulfill all the requested algebraic and differential identities.


\section{Twisted duality - Formulation without prepotentials}
\label{NoPrep}
\setcounter{equation}{0}

\subsection{Equations of motion and twisted self-duality}

Out of the graviton set (\ref{GravSet}), one can take either $h_{ij}$, $ \Phi_{ijk} $ or  $\pi^{ijk}$ as independent variables in the variational principle. The equations of motion obtained by varying the action with respect to these variables, which mutually determine each other up to gauge symmetries, are equivalent.  A similar property  also holds true for the  variables $T_{ijk}$, $ P_{ijkl} $ and $ \pi^{ij}$of the dual graviton set (\ref{DualGravSet}).  

It is instructive to verify the equivalence explicitly.  We start with the second set.  Varying the action with respect to $ \pi^{ij}$ yields the equation
\be
\dot{h}_{mn} = 2 \pi_{mn} - \frac{2}{3} \pi \delta_{mn} + n_{m,n} + n_{n,m} \label{HEOMh}
\ee
($\Leftrightarrow K_{mn} = - \pi_{mn} + \frac{\pi}{3} \delta_{mn}$).  This equation, although gauge invariant itself,  involves the Lagrange multipliers $n_m$, which are not among the phase space variables. To get a phase space expression and exhibit more explicitly the gauge invariance of (\ref{HEOMh}), one gets rid of the shift by taking the curvature of $\dot{h}_{mn}$.  No physical information is lost by doing so. 

This yields an equation of the form $R_{ijmn}[\dot{h}] =  \cdots$ where the right-hand side is obtained by acting with the operator that produces the Riemann tensor on the right-hand side of   (\ref{HEOMh}), eliminating thereby the shift.
But this is in fact the equation that one obtains by varying with respect to the prepotential $ P_{ijkl} $. One can thus say that varying the action with respect to the prepotential yields the gauge-invariant content of the equation (\ref{HEOMh}).  

The same is true if one considers the equation of motion following from extremization with respect to $T_{ijk}$, which is the Hamiltonian equation $\dot{\pi}^{ijk} = \cdots$, where the lapses $m_j$ of the Curtright theory appear differentiated twice in the right-hand side.
It turns out that one can also view the equation obtained by varying with respect to the superpotential $ P_{ijkl} $ as the gauge invariant content of this equation, obtained by projecting out $m_j$ by acting on it with the appropriate first-order differential operator. 

Identical features hold for the equations of motion associated with the set $h_{ij}$, $ \Phi_{ijk} $ or  $\pi^{ijk}$, where one finds that varying the action with respect to the prepotential $ \Phi_{ijk} $ extracts again the gauge invariant content of the equations obtained by varying with respect to $h_{ij}$ or $\pi^{ijk}$.

One can also easily verify that the equations of motion are the twisted self-duality conditions.  Indeed, the Hamiltonian equation obtained by varying the action with respect to $\pi^{ij}$ (or equivalently, $P_{ijrs}$) are, as we have just seen, equivalent to $K_{in} = - \pi_{in} + \frac{\pi}{3}\delta_{in}$ and imply
$$K_{i[n, m]} = - \pi_{i[n,m]} + \frac{1}{3}\delta_{i[n} \pi_{,m]}
$$
Multiplying the left-hand side by $(1/2)\epsilon^{rsmn}$ and summing over $(m, n)$ yields the magnetic field ${\mathcal B}_{rsi}[h]$ as it follows from the Gauss-Codazzi equations $R_{0ijk} =  \partial_j K_{ik} - \partial_k  K_{ij}$ and the definition of ${\mathcal B}_{rsi}[h]$.  Using the relation (\ref{TSchoutenP}), one sees that the similar operation applied to the right-hand side produces (minus) the electric field ${\mathcal E}_{rsi}[T]$.  Hence, the equations of motion obtained by varying with respect to the momentum $\pi^{ij}$  yield the first half of the twisted self-duality conditions,
\be
{\mathcal B}_{rsi}[h] = -{\mathcal E}_{rsi}[T]
\ee
These equations can alternatively be obtained through one integration from the variational equations derived by extremizing the action with respect to the prepotential $P_{ijrs}$, the shift components appearing then as integrating functions.  Similarly, one gets the second half of the twisted self-duality conditions
\be
{\mathcal B}_{ijrs}[T] = {\mathcal E}_{ijrs}[h]
\ee
by varying with respect to $\pi^{ijk}$ (or equivalently, the prepotential $\Phi_{ijk}$).

The comparison with the two-potential formulation of $p$-form gauge fields is again useful.  There also, one really obtains from the variational principle the ``curl" of the twisted self-duality equations.  One then integrates these variational equations and brings in an $A_0$ -- which is not really present -- to rewrite the result as twisted self-duality conditions in terms of time-space components of the curvatures.

\subsection{Variational principle with both graviton and its dual}

The dynamical description of the system must involve one variable from the ``graviton set" (\ref{GravSet}) and one variable from the conjugate ``dual graviton set" (\ref{DualGravSet}).  For instance, the Hamiltonian formulation of the Pauli-Fierz action involves ($h_{ij}$, $\pi^{ij}$), the Hamiltonian formulation of the Curtright theory involves ($T_{ijk}$, $\pi^{ijk}$), while  the prepotential formulation involves ($\Phi_{ijk}$, $P^{ijmn}$).   But one can also take ($h_{ij}$,$T_{ijk}$), or ($\pi^{ij}$, $\pi^{ijk}$). 

Adopting the system ($h_{ij}$,$T_{ijk}$) is of interest since these fields are the original graviton and dual graviton fields with a more direct  interpretation.  The analysis proceeds exactly along the lines of \cite{Bunster:2013tc}. One expresses the prepotentials in terms of ($h_{ij}$,$T_{ijk}$) by inverting the formulas giving these fields in terms of the prepotentials, which one can do when ($h_{ij}$,$T_{ijk}$) fulfill their respective Hamiltonian constraints.  The expressions are given in Appendix \ref{AppD}.
One then inserts these expressions inside the action, yielding for the Hamiltonian kinetic term
\be
 S_{K}[h,T]=\int dt\;d^4x\left[-\partial_{m}\partial_{p}\partial^{i}\Delta^{-1}\epsilon_{anij}T^{apj}-\epsilon_{amij}\partial^{i}T_{n}^{\ aj}\right]\dot{h}^{mn}\label{ActionDualMan}
\ee
This term is nonlocal in space, just as in four dimensions \cite{Bunster:2013tc}. It is, however, local in time. One verifies straightforwardly that it is invariant under the gauge symmetries of both $h_{ij}$ and $T_{ijk}$. 

The complete action is given by the sum 
\be
S[h,T, n, m_j] = S_K - \int dt d^4x \left[{\mathcal H} + n C+ m_j \Gamma^j \right]
\ee
where the Hamiltonian constraints $C= 0$, $\Gamma^j = 0$ for ($h_{ij}$,$T_{ijk}$) have been implemented by re-inserting the Lagrange multipliers $n$ and $m_j$.  The Hamiltonian density ${\mathcal H}$ is the sum of the potential energy densities of the Pauli-Fierz field  $h_{ij}$ and its conjugate $T_{ijk}$. 

The situation is similar to ordinary electromagnetism where one can eliminate the vector potential $\vec{A}$ in favor of the magnetic field $\vec{B}=\nabla \times \vec{A}$ in the action  $\int dx^0 d^3x \left( \vec{E} \cdot \dot{\vec{A}} - \mathcal{H} - A_0 (\nabla  \cdot \vec{E})\right)$ as $\vec{A}=\frac{-1}{\nabla^{2}} \nabla \times \vec{B} + gradient$.  The gradient term can be absorbed in a redefinition of the Lagrange multiplier $A_0$.  One then has an action expressed purely in terms of electric and magnetic fields, each subject to Gauss'law.  One must thus add $\nabla\cdot{B} = 0$, with its own Lagrange multiplier. The kinetic term is gauge invariant but not local in space.


\section{Conclusions}
\label{Conclusions}
\setcounter{equation}{0}

In this paper, we have shown how to reformulate linearized gravity in 5 space-time dimensions in a manner that keeps the graviton and its dual on an equal footing in the variational principle.  

We have found all the features numbered (i)-(iv) in the Introduction appearing again in the analysis, with ``manifest duality invariance" replaced by ``manifest" twisted duality. Once more, one cannot but be amazed to witness how the involved structure of the original form of the theory reveals an altogether non evident structure, if the duality principle is the torch illuminating the search.

Although we have treated explicitly only the five-dimensional case, higher dimensions can be discussed along exactly the same lines.  The dual graviton is described in the canonical formalism by a field $T_{i_1 \cdots i_{D-3} j}$ and its conjugate momentum $\pi^{i_1 \cdots i_{D-3} j}$, both of mixed symmetry $(D-3,1)$.  The ``graviton set" is now given by
$$ h_{ij}, \; \; \; \Phi_{i_1 \cdots \i_{D-3} j}, \; \; \; \pi^{i_1 \cdots i_{D-3} j} $$
where $\Phi_{i_1 \cdots \i_{D-3} j}$ is the corresponding prepotential, also of $(D-3,1)$ mixed symmetry.  Similarly, the``dual graviton set"  is now given by 
$$ T_{i_1 \cdots i_{D-3} j}, \; \; \; P_{i_1 \cdots \i_{D-3} j_1 \cdots j_{d-3}}, \; \; \; \pi^{ij} $$
where $P_{i_1 \cdots \i_{D-3} j_1 \cdots \j_{d-3}}$ is the corresponding prepotential of mixed symmetry $(D-3, D-3)$.  These two sets are not only dual to each other, but also canonically conjugate.

The spatial curvature of the dual field $T_{i_1 \cdots i_{D-3} j}$ (or of any tensor with the same mixed symmetry) is completely captured by its Ricci tensor in $d=D-1$ spatial dimensions. The Weyl tensor indeed vanishes. Similarly, the curvature of the prepotential $P_{i_1 \cdots \i_{D-3} j_1 \cdots \j_{d-3}}$ is completely determined by its multiple trace $R_{ij}[P] =R_{i k_1\cdots k_{D-3} j}^{\; \; \; \; \; \; \; \; \; \; \; \; \; \; \; \; \; \; \; \; k_1  \cdots k_{d-3}}[P]$.

We have  seen that the demand of locality  of the formalism requires the introduction of prepotentials.  These prepotentials not only guarantee locality, but also make the transition from the standard Pauli-Fierz formulation to the dual Curtright formulation more transparent, since the prepotentials are at the same time the potentials for one field and for the conjugate momentum to its dual. The need to describe gravity by variables different from the usual metric variables has arisen recently in various different contexts.  One context  which is kinematically rather similar to the present one - although the dynamics are different - is investigated in \cite{Bergshoeff:2012yz}. 

One can eliminate the prepotentials in order to express the action in terms of the graviton field $h_{ij}$ and its dual $T_{ijk}$.  However, the resulting action (\ref{ActionDualMan}) is not local in space.  It is worthwhile noting that this action has  furthermore a structure rather different from the one of the actions proposed to exhibit the hidden symmetries.  Besides being non local in space, it is of first order and linear in the time derivatives.  This is indispensable in order to avoid double-counting of the degrees of freedom, since the naive covariant second order action given by the sum of the Pauli-Fierz action plus the Curtright action without further constraint, would describe two independent spin 2 fields rather than just one.

Our work can be extended in various directions. First, one can consider  the supersymmetric formulation of the theory ($D=5, N=1$) along the lines of  \cite{Bunster:2012jp}.  Second,  as a first step to introducing interactions, it would be of interest to investigate the $D \geq 5$ theory with a cosmological constant around de Sitter space, generalizing thereby the $D=4$ treatment of \cite{Julia:2005ze}.  Finally, the full interacting theory should be understood in this duality light, a necessary step to understand the full implications of duality, in particular at the quantum level where it has been argued to play a significant role for  finiteness \cite{Kallosh:2011dp,Gunaydin:2013pma}.  We plan to return to these questions in the future.


\section*{Acknowledgments} 
C.B. and M.H.  thank  the Alexander von Humboldt Foundation for Humboldt Research Awards.  The work of M.H.and S.H. is partially supported by the ERC through the ``SyDuGraM" Advanced Grant, by IISN - Belgium (conventions 4.4511.06 and 4.4514.08) and by the ``Communaut\'e Fran\c{c}aise de Belgique" through the ARC program.  The Centro de Estudios Cient\'{\i}ficos (CECS) is funded by the Chilean Government through the Centers of Excellence Base Financing Program of Conicyt.   

\break

\noindent
{\bf \Large{Appendices}}

\appendix

\section{Weyl tensors}
\label{Weyl01}
\subsection{Weyl tensor of the Pauli-Fierz field}
The Riemann tensor of the Pauli-Fierz field $h_{\mu \nu}$ can be decomposed into $SO(D-1,1)$ irreducible components as ($D \geq 3$)
$$
R_{\al \bet \ga \de} = W_{\al \bet \ga \de} + \left(S_{\al \ga} \eta_{\bet \de} - S_{\al \de} \eta_{\bet \ga} - S_{\bet \ga} \eta_{\al \de} + S_{\bet \de} \eta_{\al \ga} \right),
$$
where the ``Weyl tensor" $W_{\al \bet \ga \de}$ fulfills 
\bea
&& W_{\al \bet \ga \de} = - W_{\bet \al \ga \de}, \; \; \; W_{\al \bet \ga \de} = -W_{\al \bet \de \ga}, \; \; \; W_{\al \bet \ga \de} = W_{\ga \de \al \bet }, \nonumber \\
&&  W_{\al \bet \ga \de} + W_{\al \ga \de \bet} + W_{\al \de \bet \ga} = 0, \nonumber 
\eea
and is traceless,
$$
W^\al_{\; \;  \; \bet \al \de} = 0.
$$
The ``Schouten tensor" $S_{\al \bet}$  is symmetric and explicitly given by
$$
S_{\al \bet} = \frac{1}{D-2} \left( R_{\al \bet} - \frac{R}{2(D-1)} \eta_{\al \bet} \right).
$$
It can be rewritten as
$$
S_{\al \bet} = \frac{1}{D-2} \left( U_{\al \bet} + \frac{(D-2)R}{2D(D-1)}  \eta_{\al \bet} \right).
$$
in terms of its irreducible components $U_{\al \bet} \equiv R_{\al \bet} - \frac{1}{D} R \eta_{\al \bet}$ ($U_{\al \bet} g^{\al \bet} = 0$) and $R$.

The Riemann tensor and the Weyl tensor coincide on-shell.  In $D$ dimensions, the Weyl tensor has 
$$ \frac{(D+2)(D+1)D (D-3)}{12}$$
components.  These are algebraically unconstrained by the equations of motion.  In five dimensions, the Weyl tensor of the Pauli-Fierz field has therefore $35$ independent components.

Under Weyl rescaling, 
\begin{equation}
\delta_C h_{\mu \nu} = 2 \xi \eta_{\mu \nu}
\end{equation}
the Weyl tensor is invariant.

\subsection{Weyl tensor of the Curtright field}
\setcounter{equation}{0}
\label{WeylT}

In $D \geq 4$ dimensions, one may similarly decompose the Riemann tensor $E_{\al \beta \gamma \rho \si}$ of the Curtright field into a traceless (``Weyl") part plus terms containing the Ricci tensor and the vector curvature.  One finds explicitly
\begin{eqnarray}
\hspace{-1cm} E_{\al \beta \gamma \rho \si} &=& W_{\al \beta \gamma \rho \si}  
 - \frac{1}{D-3} \Big[ ( \eta_{\al \rho} S_{\beta \gamma \si}  - \eta _{\al \si} S_{\beta \gamma \rho})   
\nonumber \\ && \hspace{1.5cm} + (\eta_{\beta \rho} S_{ \gamma \al \si} - \eta _{\beta \si} S_{ \gamma \al \rho}) + (\eta_{\gamma \rho} S_{\al \beta \si} - \eta _{\gamma \si} S_{\al \beta  \rho}) \Big]  
\end{eqnarray}
where the ``Schouten tensor" $S_{\al \bet \rho}$ is given by
$$
S_{\al \bet \rho} = E_{\al \bet \rho} + \frac{1}{2(D-2)}(\eta_{\al \rho} E_\beta - \eta_{\beta \rho} E_\al),
$$
or equivalently
$$
S_{\al \bet \rho} = U_{\al \bet \rho} - \frac{D-3}{2(D-1)(D-2)}(\eta_{\al \rho} E_\beta - \eta_{\beta \rho} E_\al),
$$
where $U_{\al \beta \rho}$ is traceless, $U_{\al \beta \rho} \eta^{\beta \rho} = 0$.  Again, the Riemann and Weyl tensors coincide on-shell. The Weyl tensor is algebraically non-constrained by the equations of motion and has
$$
\frac{(D+2)(D+1)D(D-1)(D-4)}{24}
$$ independent components.  In five dimensions,  the Weyl tensor of the Curtright field has therefore $35$ independent components.

Under Weyl rescalings of the Curtright field,
\begin{equation}
\delta_C T_{\mu\nu \rho} = \eta_{\mu \rho} \xi_\nu - \eta_{\nu \rho} \xi_\mu
\end{equation}
the Weyl tensor is invariant, while the Schouten tensor transforms as
\be
\delta_C S_{\alpha_1 \al_2 \beta} = (D-3) \left(\partial_{\al_1} \partial_\beta \xi_{\al_2} - \partial_{\al_2} \partial_\beta \xi_{\al_1}\right)
\ee
(a formula that can be read as $\delta_C S = (D-3)d^2 \xi$ in the terminology of \cite{DVH}). One defines the Cotton tensor as 
\be
C_{\alpha_1 \al_2 \beta_1 \beta_2} = \partial_{\beta_2} S_{\al_1 \al_2 \beta_1} - \partial_{\beta_1} S_{\al_1 \al_2 \beta_2} .
\ee
This tensor is not of irreducible Young symmetry type since it contains both components of  $$(2,2) \equiv \yng (2,2)$$  and $$ (3,1) \equiv \yng(2,1,1)$$ Young symmetry types.  Explicitly, with
\be
M_{\alpha_1 \al_2 \beta_1 \beta_2} = \partial_{\beta_2} S_{\al_1 \al_2 \beta_1} - \partial_{\beta_1} S_{\al_1 \al_2 \beta_2} + \partial_{\al_2} S_{\beta_1 \beta_2 \al_1} - \partial_{\al_1} S_{\beta_1 \beta_2 \al_2}
\ee
(of $(2,2)$-type) and
\be
N_{\alpha_1 \al_2 \beta_1 \beta_2} =   \partial_{\beta_1} S_{\al_1 \al_2 \beta_2} + \partial_{\al_2} S_{\beta_1 \al_1 \beta_2} + \partial_{\al_1} S_{\al_2 \beta_1 \beta_2}
\ee
(of $(3,1)$-type), one has
\be
C_{\alpha_1 \al_2 \beta_1 \beta_2} = \frac{1}{2} M_{\alpha_1 \al_2 \beta_1 \beta_2} - \frac{1}{2} (N_{\alpha_1 \al_2 \beta_1 \beta_2} - N_{\alpha_1 \al_2 \beta_2 \beta_1}).
\ee
There is no component of
$$ (4) \equiv \yng(1,1,1,1)$$ Young symmetry type because
\be
C_{[\alpha_1 \al_2 \beta_1 \beta_2]} = 0. \label{zeroDT}
\ee

The Cotton tensor is traceless because of the Bianchi identity, 
\begin{equation}
C_{\alpha_1 \al_2 \beta_1 \beta_2} \eta^{\al_2 \beta_2} = 0
\ee
and clearly obeys
\be
C_{\alpha_1 \al_2 [\beta_1 \beta_2, \beta_3]} = 0. \label{DivVan}
\ee
It is also invariant under Weyl rescalings, 
\be
\delta_C C_{\alpha_1 \al_2 \beta_1 \beta_2} = 0
\ee
because $\delta_C C  \sim d^3 \xi = 0$.  Finally, when $D \not=4$, it can be expressed in terms of  the divergence of the Weyl tensor through the Bianchi identity,
\be
\partial^\mu W_{\mu \al \beta \rho \sigma} - \frac{D-4}{D-3} C_{ \al \beta \rho \sigma} = 0
\ee
and so does not provide an independent Weyl invariant.  However, when $D= 4$, the Weyl tensor vanishes and the Cotton tensor $C_{\alpha_1 \al_2 \beta_1 \beta_2}$ is an independent Weyl invariant.  This is in complete analogy with the well-known situation found for the curvature of a $(1,1)$- tensor in 3 dimensions, where the Weyl tensor identically vanishes and the Cotton tensor provides a (complete) set of independent Weyl invariants. 

It is useful to introduce the co-Cotton tensor 
\be
D^{\al_1 \cdots \al_{D-2} \beta_1 \beta_2} = \frac{1}{2} \epsilon^{\al_1 \cdots \al_{D-2}}_{\ \ \ \  \ \ \ \  \ \ \ga_1 \ga_2} C^{\beta_1 \beta_2 \ga_1 \ga_2}
\ee
This is a tensor of irreducible $(D-2,2)$ Young type because the Cotton tensor is traceless,
\be D^{[\al_1 \cdots \al_{D-2} \beta_1] \beta_2} = 0, \ee  which has furthermore zero double-trace because of (\ref{zeroDT}),
\be D^{\al_1 \al_2 \cdots \al_{D-2} \beta_1 \beta_2} \eta_{\al_1 \beta_1} \eta_{\al_2 \beta_2} = 0. \ee  Its divergence also vanishes
\be
\partial_{\al_1}D^{\al_1 \al_2 \cdots \al_{D-2} \beta_1 \beta_2} = 0
\ee
because of (\ref{DivVan}).  Note that this relation implies
\be
\partial_{\beta_1}D^{\al_1 \al_2 \cdots \al_{D-2} \beta_1 \beta_2} = 0
\ee

\subsection{Comments on a $(2,2)$ field in 4 dimensions}
\label{Cotton22}

Since this is needed for the understanding of the prepotential for the momenta, we now consider a $(2,2)$ field $P_{ijrs}$ in four dimensions with gauge symmetries
\be
\delta_1 P_{ijrs} = \chi_{rs[i,j]} + \chi_{ij[r,s]}  \label{gaugeP}
\ee
and
\be
\delta_C P_{ijrs} = \frac{1}{4} [\delta_{ir} \delta_{js} - \delta_{is} \delta_{jr}] \xi \label{WeylP}
\ee (Weyl rescalings in a space of Euclidean signature).   The curvature $R_{ijkrst}$ invariant under the gauge transformations (\ref{gaugeP}) (which may be written as $\delta_1 P = d \chi$ \cite{DVH}) is a $(3,3)$ tensor given by
\be
R_{ijkrst}= 18 \partial_{[i}P_{jk][rs,t]}
\ee
($R = d^2 P$).  In 4 dimensions, this tensor is completely determined by its double-trace $R_{ir} \equiv R_{ijkrst} \delta^{js} \delta^{kt} $ as follows,
\begin{eqnarray}
&&R_{ijkrst} = \frac{1}{2} \left[ \left(\delta_{ir} \delta_{js} - \delta_{is} \delta_{jr} \right) S_{kt} + \left(\delta_{is} \delta_{jt} - \delta_{it} \delta_{js} \right) S_{kr} + \left(\delta_{it} \delta_{jr} - \delta_{ir} \delta_{jt} \right) S_{ks}\right] \nonumber \\
&& \hspace{.1cm} + \frac{1}{2} \left[ \left(\delta_{jr} \delta_{ks} - \delta_{js} \delta_{kr} \right) S_{it} + \left(\delta_{js} \delta_{kt} - \delta_{jt} \delta_{ks} \right) S_{ir} + \left(\delta_{jt} \delta_{kr} - \delta_{jr} \delta_{kt} \right) S_{is}\right] \nonumber \\
&& \hspace{.1cm} + \frac{1}{2} \left[ \left(\delta_{kr} \delta_{is} - \delta_{ks} \delta_{ir} \right) S_{jt} + \left(\delta_{ks} \delta_{it} - \delta_{kt} \delta_{is} \right) S_{jr} + \left(\delta_{kt} \delta_{ir} - \delta_{kr} \delta_{it} \right) S_{js}\right]
\end{eqnarray}
where the coresponding ``Schouten tensor" $S_{ij}$ is equal to 
\be 
S_{ij} = R_{ij} - \frac{2}{9} \delta_{ij} R.
\ee
Under Weyl rescalings (\ref{WeylP}) of $P_{ijrs}$, the Schouten tensor transforms as
\be
\delta_C S_{ij} = \partial_i \partial_j \xi.
\ee

One defines the Cotton tensor $C_{ijk}$ as
\be C_{ijk} = \partial_{[i} S_{j]k}.
\ee
This is a $(2,1)$-tensor because $S_{jk}$ is symmetric.  The Cotton tensor is invariant under Weyl rescalings.  It is also traceless as a consequence of the Bianchi identity $R_{ijk[rst,u]}= 0$ and  fulfills $\partial_{[l} C_{ij]k} = 0$.  The tracelessness of the Cotton tensor is just the contracted Bianchi identity
\be
\partial_i G^{ij} = 0
\ee 
for the Einstein tensor 
\be
G^{ij} = R^{ij} - \frac{1}{3} \delta^{ij} R
\ee
of $P_{ijrs}$.  

 The co-Cotton tensor is defined as
\be
D^{rsk} = \frac{1}{2} \epsilon^{rsij} C_{ij}^{\ \  k}
\ee
It is a $(2,1)$-tensor that has the properties
\be
\delta_C D^{ijr} = 0, \; \; \; D^{ij}_{\ \ j} = 0, \; \; \partial_i D^{ijk} = 0, \; \; \partial_k D^{ijk} = 0.
\ee

\section{Spacetime decomposition of the curvature of the Curtright field}
\setcounter{equation}{0}
\label{AppDecomp}

\subsection{Decomposition in space and time of the Einstein tensor}
\subsubsection{Formulas}
We recall the formulas
$$R_{0i0j} =  \partial_0 K_{ij} - \frac{1}{2} \partial_i \partial_j h_{00}$$
as well as
$$R_{0ijk} =  \partial_j K_{ik} - \partial_k  K_{ij}, \; ^{(5)} \! R_{ijkl} = \, ^{(4)} \! R_{ijkl} $$
(linearized Gauss-Codazzi equations).  This implies
$$G_{00} =  \frac12 \; ^{(4)} \! R, \; \; R_{0i} = - \partial^m (K_{im} - \delta_{im} K)$$
and $$ ^{(5)} \! R_{ij} = - \partial_0 K_{ij} + \frac12 \, \partial_i \partial_j h_{00} +  \, ^{(4)} \!R_{ij} . $$

\subsubsection{Electric and magnetic fields}
The 35 independent components of the Weyl tensor of the Pauli-Fierz field can be decomposed into the 19 independent ``electric components" $^{(5)} \!W_{ijrs}$ and the 16 independent ``magnetic components" $W_{0irs}$.  The components $W_{0i0r}$ with two time indices are not independent from the electric components  $^{(5)} \!W_{ijrs}$ since one has $- W_{0i0r} + ^{(5)} \!W_{kisr} \delta^{ks} = 0$. [General properties of decompositions of tensors in $D$ spacetime dmensions into ``electric" and ``magnetic" components are studied in \cite{Senovilla:1999xz}.]

There are only 19 independent components among the 20 components $^{(5)} \!W_{ijrs}$ (which form a spatial $\yng(2,2)$-tensor) because the $^{(5)} \!W_{ijrs}$ fulfill the double tracelessness condition,
\begin{equation}
^{(5)} \!W_{ijrs} \delta^{js} \delta^{ir} = W_{i0r0} \delta^{ir} = W_{0000} = 0
\end{equation}
since the Weyl tensor $W_{\al \bet  \rho \si}$ is traceless. 

Similarly, there are only 16 independent components among the 20 components $W_{0irs}$ (which spatially  transform in the $\yng(2,1)$-representation since $W_{0[irs]} =0$) because these are subject to the 4 tracelessness conditions
\begin{equation}
W_{0irs} \delta^{is} = 0
\end{equation}
as a consequence of the tracelessness of $W_{\al \bet  \rho \si}$.

The electric components $^{(5)} \!W_{ijrs}$ of the Weyl tensor are equal on-shell to the components $R_{ijrs}$ of the Riemann tensor  ($^{(5)} \! R_{ijrs} = ^{(4)} \! R_{ijrs} \equiv R_{ijrs}$).  We define the electric field ${\mathcal E}_{ijrs}[h] $ to be the double spatial dual of the Riemann tensor,
\begin{equation}
{\mathcal E}_{ijrs}[h] = \frac{1}{4} \epsilon_{ijmn} R^{mnpq} \epsilon_{pqrs}.
\end{equation}
Explicitly one gets,
\begin{equation}{\mathcal E}_{ijrs}[h] =   R_{ijrs} - \delta_{ir} R_{js}+  \delta_{is} R_{jr} + \delta_{jr} R_{is} -  \delta_{js}  R_{ir}  + \frac12 (\delta_{ir} \delta_{js} - \delta_{is} \delta_{jr})R
\end{equation} where we recall that $R_{js}$ is the Ricci tensor of the spatial metric, namely, $R_{js} \equiv ^{(4)} \! \! \! R_{js}$, and $R \equiv ^{(4)} \! \! \! R$.
The electric field is a spatial tensor of type $\yng(2,2)$.  One has ${\mathcal E}_{ir}[h] = - R_{ir} + \frac12 \delta_{ir} R$, and  ${\mathcal E}[h] =  R$ where ${\mathcal E}_{ir}[h]$ is the ``Ricci tensor" ${\mathcal E}_{ijrs}[h] \delta^{js}$ of the electric field and ${\mathcal E}$ its double-trace ${\mathcal E}_{ijrs}[h] \delta^{js} \delta^{ir}$, so that one can express $R_{ijrs}$ in terms of $ {\mathcal E}_{ijrs}[h]$ through a formula that takes of course exactly the same form (since the operation of taking the double-dual yields the identity when squared),
$$R_{ijrs} =   {\mathcal E}_{ijrs} - \delta_{ir} {\mathcal E}_{js}+  \delta_{is} {\mathcal E}_{jr} + \delta_{jr} {\mathcal E}_{is} -  \delta_{js}  {\mathcal E}_{ir}  + \frac12 (\delta_{ir} \delta_{js} - \delta_{is} \delta_{jr}){\mathcal E}. $$
Knowledge of the electric field ${\mathcal E}_{ijrs}[h] $ is equivalent to the knowledge of the spatial curvature $R_{ijrs}$ and vice-versa. The linear operator that relates the two is not only invertible but also idempotent.  For the purpose of discussing the $(4+1)$-form of the twisted self-duality equations (\ref{122}), it turns out that the electric field is the natural object to consider as it is the electric field that appears in the formulas.

All 20 components of ${\mathcal E}_{ijrs}[h] $ are independent off-shell, but only 19 of them are independent on-shell.  The equations of motion imply indeed that ${\mathcal E}_{ijrs}[T] $ is doubly traceless, 
${\mathcal E}_{ijrs}[h] \delta^{js} \delta^{ir} = R = 0
$.
 This is in fact just the $G_{00}=0$ constraint equation.

In the same way, the magnetic components $W_{0irs}$ of the Weyl tensor are equal on-shell to the components $R_{0irs}$ of the Riemann tensor.  We define the magnetic field ${\mathcal B}_{rsi}[h] $ to be 
\begin{equation}{\mathcal B}_{rsi}[h] =  \frac{1}{2} R_{0imn} \, \epsilon_{rs}^{\; \; \; \;  mn}
\end{equation}
The magnetic field completely captures all the magnetic components of the Weyl tensor on-shell.  Off-shell, the magnetic field does not fulfill the condition ${\mathcal B}_{[irs]}[h] =0$ and so does not transform in an irreducible representation.   Since the traces ${\mathcal B}_{[irs]}[h] \delta^{is}$  identically vanish, it possesses 20 independent off-shell components. The 4 independent conditions ${\mathcal B}_{[irs]}[h] =0$  arise on-shell and are equivalent to the constraint equations $R_{0j} = 0$.  They imply that on-shell, the magnetic field transforms in the irreducible representation $\yng(2,1)$ and possesses therefore 16 independent components (given the tracelessness conditions). 

Both the electric and the magnetic fields are transverse,
\begin{equation}
 \partial^m {\mathcal E}_{mnrs}[h] = 0 , \; \; \;  \partial^m {\mathcal E}_{mn} [h] = 0 , \; \; \; 
 \partial^m {\mathcal B}_{mnr}[h] = 0.
\end{equation}

\subsection{Invariant velocities for the Curtright field}
The velocities $\dot{T}_{ijk}$ of the Curtright field transform under gauge transformations with the time derivatives of the gauge parameters $\si_{\al \beta}$, $\al_{\al \beta}$. For that reason, these ordinary velocities, given on an initial hypersurface $x^0 = 0$ (say), are not invariant under gauge transformations $\si_{\al \beta}$, $\al_{\al \beta}$ that reduce to the identity on that hypersurface,$$\si_{\al \beta}(x^0=0,x^k)=0, \; \; \; \; \al_{\al \beta}(x^0=0, x^k) = 0$$ since one may have $\dot{\si}_{\al \beta}(x^0=0,x^k)\not=0$ or $\dot{\al}_{\al \beta}(x^0=0, x^k) \not= 0$.   As pointed out by Dirac, it is useful to introduce  variables that have the desirable property of transforming only with the gauge parameters taken at the same time, i.e., whose gauge variations do not involve the time derivatives of the gauge parameters \cite{Dirac}.  For the Curtright field, these ``invariant velocities" $V_{ijk}$ are given by
\begin{equation}
V_{ijk} = \dot{T}_{ijk} + \partial_i T_{j0k} - \partial_j T_{i0k} - \partial_k T_{ij0}.
\end{equation}
They transform as
\begin{equation}
\delta V_{ijk} = \partial_k\Big( \partial_j \left(\si_{0i} - 3 \al_{0i} \right) -   \partial_i \left(\si_{0j} - 3 \al_{0j} \right) \Big) \label{GaugeInvVel}
\end{equation}
and play a role analogous to that of the extrinsic curvature for the symmetric tensor $\yng(2)$.  The  invariant velocities $V_{ijk}$  are the components of a tensor of type $\yng(2,1)$.

\subsection{Explicit formulas}
We now decompose the components of the curvature tensor $E_{\al \beta \gamma \mu \nu}$ into space and time.
\subsubsection{The components $E_{0ij0k}$}
We start with the components with two time indices. One finds
\begin{equation}
E_{0ij0k} =- \partial_0V_{ijk} - \partial_k \left(\partial_i T_{0j0} - \partial_j T_{0i0} \right) ,
\end{equation}
an expression which is easily checked to be gauge invariant.

\subsubsection{The components $E_{0ijkm}$ and $E_{ijk0m}$}
Turning to the components $E_{0ijkm}$, we get
\begin{equation}
E_{0ijkm} = \partial_m V_{ijk} -  \partial_k V_{ijm}
\end{equation}
and
\begin{equation}
E_{ijk0m} =  -  3 \partial_{[k} V_{ij]m} = 3 E_{0[ijk]m}
\end{equation}
(the last expression in agreement with $E_{[0ijk]m} = 0$).

\subsubsection{The components $^{(5)} \! E_{ijkmn}$}
{}Finally, for the components $^{(5)} \! E_{ijkmn}$, we have of course
\begin{equation}
^{(5)} \! E_{ijkmn} = E_{ijkmn}.
\end{equation}

\subsection{Decomposition of the Ricci tensor $E_{\al \bet \mu}$}
Straightforward computations yield for the non-vanishing components of $E_{\al \bet \mu}$,
\begin{eqnarray}
 E_{0i0} &=& - \partial_0 V_{ik}^{\; \; \; \; k} - \partial^k (\partial_i T_{0k0} - \partial_k T_{0i0})\\
 E_{0ij} &=& \partial^k V_{ikj} - \partial_j V_{ik}^{\; \; \; \; k}\\
 E_{ij0} &=& -\partial_i V_{jk}^{\; \; \; \; k} - \partial_j V_{ki}^{\; \; \; \; k} - \partial^k V_{ijk}\\
 ^{(5)} \! E_{ijk} &= &  - \partial_0 V_{ijk} - \partial_k (\partial_i T_{0j0} - \partial_j T_{0i0}) + E_{ijk}
\end{eqnarray}

\subsection{Electric and magnetic fields}
The 35 independent components of the Weyl tensor of the Curtright field can be decomposed into the 16 independent ``electric components" $^{(5)} \!W_{ijkrs}$ and the 19 independent ``magnetic components" $W_{0ijrs}$.  The components $W_{0ij0r}$ with two time indices are not independent from the electric components since one has $- W_{0ij0r} + ^{(5)} \!W_{kijsr} \delta^{ks} = 0$.  Furthermore, the components with one time index $W_{ijk0r}$ are not independent from $W_{0ijkr}$ since one has
\begin{equation}
W_{ijk0r} = 3 W_{0[ijk]r}.
\end{equation}

There are only 16 independent components among the 20 components $^{(5)} \!W_{ijkrs}$ (which form a spatial $\yng(2,1,1)$-tensor) because the $^{(5)} \!W_{ijkrs}$ fulfill the double tracelessness condition,
\begin{equation}
^{(5)} \!W_{ijkrs} \delta^{ks} \delta^{jr} = W_{ij0r0} \delta^{jr} = W_{i0000} = 0
\end{equation}
since the Weyl tensor $W_{\al \bet \ga \rho \si}$ is traceless and of $\yng(2,1,1)$-type. 

Similarly, there are only 19 independent components among the 36 components $W_{0ijrs}$ (which spatially  transform a priori in the $\yng(1,1) \otimes \yng(1,1) = \yng(2,2) \oplus \yng(2,1,1) \oplus \yng(1,1,1,1)$) because these are subject to the 16 tracelessness conditions
\begin{equation}
W_{0ijrs} \delta^{js} = 0
\end{equation}
(as a consequence of the tracelessness of $W_{\al \bet \ga \rho \si}$) and the fully antisymmetric condition
\begin{equation}
W_{0[ijrs]} =0
\end{equation}
(as a consequence of $3 W_{0[ijrs]} = - W_{[ijkr]0} = 0$), which eliminates the $\yng(1,1,1,1)$-representation.

The electric components $^{(5)} \!W_{ijkrs}$ of the Weyl tensor are equal on-shell to the components $E_{ijkrs}$ of the Riemann tensor  ($^{(5)} \!E_{ijkrs} = ^{(4)} \!E_{ijkrs} \equiv E_{ijkrs}$).  Furthermore, in four dimensions, the Riemann tensor $E_{ijkrs}$ is completely equivalent to the Ricci tensor $E_{ijk}$ since the Weyl tensor vanishes ($^{(4)} \!W_{ijkrs}= 0$).  We define the electric field ${\mathcal E}_{ijr}[T] $ to be 
\begin{equation}{\mathcal E}_{ijr}[T] =  G_{ijr}
\end{equation}
(Einstein tensor of the space like $T_{ijk}$). It is a spacelike tensor of type $\yng(2,1)$ that is completely equivalent on-shell, as we have just seen, to $^{(5)} \!W_{ijkrs}$.  
Off-shell, all 20 components of ${\mathcal E}_{ijr}[T] $ are independent.  The equations of motion imply that ${\mathcal E}_{ijr}[T] $ is traceless, 
\begin{equation}{\mathcal E}_{ijr}[T] \delta^{jr} = 0
\end{equation}
since the double-trace of $^{(5)} \!W_{ijkrs}$ vanishes, yielding 16 independent on-shell components as it should.  These equations are in fact just the $G_{0i0}=0$ constraint equations.

In the same way, the magnetic components $W_{0ijrs}$ of the Weyl tensor are equal on-shell to the components $E_{0ijrs}$ of the Riemann tensor.  We define the magnetic field ${\mathcal B}_{ijrs}[T] $ to be 
\begin{equation}{\mathcal B}_{ijrs}[T] =  \frac{1}{2} E_{0ijmn} \, \epsilon_{rs}^{\; \; \; \;  mn}
\end{equation}
The magnetic field completely captures all the magnetic components of the Weyl tensor on-shell.  Off-shell, the magnetic field does not fulfill the condition ${\mathcal B}_{[ijr]s}[T] =0$ and so does not transform in an irreducible transformation.   Since it has a double-trace that is identically zero, it possesses 35 independent off-shell components. The 16 independent conditions ${\mathcal B}_{[ijr]s}[T] =0$  arise on-shell and are equivalent to the constraint equations $E_{0ij} = 0$.  They imply that on-shell, the magnetic field transforms in the irreducible representation $\yng(2,2)$ and possesses therefore 19 independent components (given the double-tracelessness condition). 

The electric and magnetic fields of $T_{\lambda \mu \rho}$ are also identically transverse,
\begin{equation}
\partial^m {\mathcal E}_{mnr}[T] = 0, \; \; \;  
\partial^r {\mathcal B}_{mnrs}[T] = 0 , \; \; \;  \partial^n {\mathcal B}_{mn} [T] = 0 , 
\end{equation}

\section{Solving the Hamiltonian constraint of the Curtright theory}
\setcounter{equation}{0}
\label{SHCCT}
The Curtright field solves the  constraints $\Gamma^{j}=0$ if and only if the tensor $t_{ijk}$ in the decomposition (\ref{DecompT}) is a solution of
\begin{equation}
\partial_{i}\partial_{k}t^{ijk}=0 \label{cst}
\end{equation}
This equation can be written as $\partial_i B^{ij} = 0$ with $B^{ij} = \partial_k t^{ijk} = - B^{ji}$ and implies therefore, by Poincar\'e lemma, $B^{ij} = \partial_k A^{ijk}$, for some completely antisymmetric tensor $A^{ijk}= A^{[ijk]}$.  Hence, $\partial_k (t^{ijk} - A^{ijk}) = 0$ so that , using Poincar\'e lemma again, we obtain
\begin{equation}
t^{ijk}= A^{ijk} + \partial_{l}N^{[kl][ij]}
\end{equation}
where the antisymmmetry on the indices of $N$ have been emphasized. After projecting onto the (2,1) symmetry of $t$, the completely antisymmetric tensor $A^{ijk}$ drops out and this relation becomes
\begin{equation}
t^{ijk}=\frac{1}{3}\partial_{l}\left[2N^{[kl][ij]}+N^{[il][kj]}-N^{[jl][ki]}\right].
\end{equation}

It is useful to dualize on the first pair of indices and redefine the dualized field by a scaling factor of -2 introduced for future convenience ($N^{[kl][ij]}=-2\epsilon^{klab}P_{[ab]}^{\ \ \ \ [ij]}$).  This yields
\begin{equation}
t^{ijk}=-\frac{2}{3}\partial_{l}\left[2\epsilon^{klab}P_{[ab]}^{\ \ \ [ij]}+\epsilon^{ilab}P_{[ab]}^{\ \ \ [kj]}-\epsilon^{jlab}P_{[ab]}^{\ \ \ [ki]}\right] \label{tPbis}
\end{equation}

At this stage, the tensor $P_{[ab][cd]}$ need not transform in an irreducible representation of the linear group and may contain components with the $(2,2)$, $(3,1)$ and $(4)$ Young symmetries, respectively. One has quite generally
\begin{equation}
P_{[ab][cd]}=R_{abcd}+(Q_{abcd}-Q_{abdc})+ A_{abcd}\label{P}
\end{equation}
where $R_{abcd} = R_{[ab][cd]}$, $R_{[abc]d} =0$ ($(2,2)$-component), $Q_{abcd}=Q_{[abc]d}$, $Q_{[abcd]}=0$ ($(3,1)$-component) and $A_{abcd}= A_{[abcd]}$ is totally antisymmetric in all its indices ($(4)$-component).  The $R$-term and the $A$-term are symmetric for the exchange of the pairs $(ab)$ with $(cd)$, while the $Q$-term is antisymmetric,
$$ R_{abcd} = R_{cdab}, \; \; A_{abcd} = A_{cdab}, \; \; Q_{abcd} - Q_{abdc} = -(Q_{cdab} - Q_{cdba}) .$$  A useful relation is $Q_{abcd} = 3 Q_{d[abc]}$.
One can express the irreducible components in terms of $P_{[ab][cd]}$ as follows,
\begin{eqnarray}
&& \hspace{-.5cm} R_{abcd} =  \frac13 (P_{abcd} + P_{cdab}) - \frac16 (P_{acdb} + P_{dbac} + P_{adbc} + P_{bcad}) \hspace{.5cm}\\
&& \hspace{-.5cm} Q_{abcd} = \frac34 (P_{[abc]d} + P_{d[abc]}) \label{QP}\\
&& \hspace{-.5cm} A_{abcd} = P_{[abcd]}
\end{eqnarray}
with $P_{abcd} \equiv P_{[ab][cd]}$.

We shall now check that one can remove the $(3,1)$ and $(4)$-components from $t^{ijk}$ through a redefinition of $u_{ij}$ and $v_{ij}$ (gauge transformation).  To that end, we observe that the tracelessness condition $t_{ik}^{\ \ k}=0$ implies
\begin{equation}
\partial_{l}\epsilon^{klab}P_{abik}=\partial_{l}\epsilon^{klab}T_{abik}=0
\end{equation}
where we have defined $T_{abik}=-T_{baik}=-T_{abki}= Q_{abik}-Q_{abki}+ A_{abik}$ (the $R$-component drops out because of $R_{[abc]d} =0$). This equation implies
\[
T_{[abk]i}=\partial_{\left[k\right.}B_{\left.ab\right]i}
\]
for some tensor $B_{abi} = - B_{bai}$.  Using the above formulas, this yields then
\[
A_{abik}=T_{[abik]}=\partial_{[i}B_{abk]}.
\]
Similarly, from (\ref{QP}), one gets
$$Q_{abcd} = \frac38 \left(3 T_{[abc]d} + T_{[dbc]a} + T_{[adc]b} + T_{[abd]c} \right) $$
and thus
\begin{eqnarray}Q_{abik}-Q_{abki} &=& \frac{3}{4}\left(T_{[abi]k}+T_{[kbi]a}+T_{[aki]b}-T_{[abk]i]}\right)
\nonumber \\
&=& \frac34 \left(\partial_{\left[i\right.}B_{\left.ab\right]k} + \partial_{\left[i\right.}B_{\left.kb\right]a} + \partial_{\left[i\right.}B_{\left.ak\right]b} - \partial_{\left[k\right.}B_{\left.ab\right]i}\right) \nonumber
\end{eqnarray}
Inserting these expressions for $A_{abik}$ and $Q_{abik}$ in terms of  $\partial_{[m} B_{cd]n}$ in (\ref{tPbis}) and (\ref{P}) shows, after some cumbersome but direct computation, that the irreducible components $A_{abik}$ and $Q_{abik}$ correspond to gauge transformation terms and hence can indeed be absorbed in a redefinition of the prepotentials $u_{ik}$ and $v_{ik}$. We leave the details to the reader.  In fact, the structure of (\ref{tPbis}) and the form of the  $Q$-component and the $A$-component imply, given that $\ep^{mnla} \partial_l \partial_a = 0$, that the $Q$ and $A$ contributions to $t_{ijk}$ can be expressed as
$\partial_i L_{jk}+ \partial_j M_{ik} + \partial_k N_{ij} $, for some tensors $ L_{jk}$,  $M_{ik}$ and $N_{ij}$.  Since $t_{ijk}$ has the Young symmetry $(2,1)$, these contributions come projected on this symmetry type, and this is precisely the most general form of a gauge transformation of the Curtright field.

We can therefore assume that the prepotential $P_{abcd}$ reduces to $R_{abcd}$, i.e., transform in the irreducible representation $(2,2)$,
\begin{equation} P_{abcd} = P_{[ab][cd]}, \; \; \; P_{[abc]d} = 0. \end{equation}

\section{Inversion Formulas}
\label{AppD}
\setcounter{equation}{0}

We give in this appendix the inversion formulas that express the prepotentials in terms of the original canonical variables. These relations hold only when the canonical variables obey the constraint equations. They involve furthermore gauge choices, since the prepotentials are determined by the canonical variables up to gauge transformations.

From  (\ref{h})-(\ref{jPhi}), one gets
\be
\Phi_{mst}=-\frac{1}{12}\left[2\epsilon_{stab}\frac{\partial^{a}}{\Delta}h_{m}^{\ b}+\epsilon_{mtab}\frac{\partial^{a}}{\Delta}h_{s}^{\ b}-\epsilon_{msab}\frac{\partial^{a}}{\Delta}h_{t}^{\ b}\right]
\ee
since
$$
\partial^{k}\left[\epsilon_{nkst}\phi_{m}^{\ st}+\epsilon_{mkst}\phi_{n}^{\ st}\right]=h_{mn}+\partial_{m}\left(8\frac{\partial^{k}}{\Delta}j_{nk}\right)+\partial_{n}\left(8\frac{\partial^{k}}{\Delta}j_{mk}\right).
$$

Similarly,  (\ref{piPhi})  yields
\begin{equation}
\phi[\pi]_{abn}=-\frac{1}{2\Delta}\pi_{abn}
\end{equation}
That this expression is correct can again be checked directly by inserting it back into (\ref{piPhi}).  One gets,
\[
-\frac{1}{2\Delta}\partial^{l}\partial^{m}\epsilon_{ijln}\epsilon_{kmab}\pi^{abn}=\pi_{ijk}-\frac{\delta_{ik}}{\Delta}\left[\partial_{j}\partial_{a}\pi^{an}_{\ \ \ n}-\Delta\pi_{jn}^{\ \ \ n}+\partial_{a}\partial_{n}\pi^{jan}\right]+
\]
\[
+\frac{\delta_{jk}}{\Delta}\left[\partial_{i}\partial_{a}\pi^{an}_{\ \ \ n}-\Delta\pi_{in}^{\ \ \ n}+\partial_{a}\partial_{n}\pi^{ian}\right]-\frac{1}{\Delta}\partial_{i}\partial_{k}\pi_{jn}^{\ \ \ n}+\frac{1}{\Delta}\partial_{k}\partial_{j}\pi_{in}^{\ \ \ n}-
\]
\[
-\frac{1}{\Delta}\partial_{k}\partial_{n}\pi_{ij}^{\ \ \ n}-\frac{1}{\Delta}\partial^{n}\partial_{j}\pi_{ink}+\frac{1}{\Delta}\partial^{n}\partial_{i}\pi_{jnk}=\pi_{ijk}+\delta\pi_{ijk}
\]
where $\delta\pi_{ijk}=-\frac{\delta_{ik}}{\Delta}\left[\partial_{j}\partial_{a}\pi^{an}_{\ \ \ n}-\Delta\pi_{jn}^{\ \ \ n}\right]+\frac{\delta_{jk}}{\Delta}\left[\partial_{i}\partial_{a}\pi^{an}_{\ \ \ n}-\Delta\pi_{in}^{\ \ \ n}\right]-\frac{1}{\Delta}\partial_{i}\partial_{k}\pi_{jn}^{\ \ \ n}+\frac{1}{\Delta}\partial_{k}\partial_{j}\pi_{in}^{\ \ \ n}$, bearing in mind that the constraints must be satisfied.  The term $\delta\pi_{ijk}$ is a gauge transformation.

In the same manner, one gets from (\ref{tP}) the following expressions for the prepotential $P_{ijkl} $ in terms of $T_{ijk}$ or $\pi^{ij}$.  
\begin{eqnarray}
P_{abcd}[T] &=&\frac{3}{16}\left[\epsilon_{abij}\frac{\partial^{i}}{\Delta}T_{cd}^{\ \ j}+\epsilon_{cdij}\frac{\partial^{i}}{\Delta}T_{ab}^{\ \ j}\right]
\nonumber \\
&&-\frac{1}{16}\left[\epsilon_{abij}\frac{\partial^{i}}{\Delta}T_{cd}^{\ \ j}+\epsilon_{cdij}\frac{\partial^{i}}{\Delta}T_{ab}^{\ \ j}+\epsilon_{caij}\frac{\partial^{i}}{\Delta}T_{bd}^{\ \ j} \right. \nonumber \\
&& \left. +\epsilon_{adij}\frac{\partial^{i}}{\Delta}T_{bc}^{\ \ j}+\epsilon_{bcij}\frac{\partial^{i}}{\Delta}T_{ad}^{\ \ j}+\epsilon_{bdij}\frac{\partial^{i}}{\Delta}T_{ca}^{\ \ j}\right]
\label{inv}
\end{eqnarray}
A straightforward but tedious computation shows that this expression, when inserted back in the relation giving $T_{rsk}$ in terms of $P_{mnrs}$ reproduces indeed $T_{rsk}$ up to a gauge transformation.

Finally, one has
\be
P_{abcd}[\pi]=-\frac{1}{4}\triangle^{-1}\left[\delta_{bd}\pi_{ac}-\delta_{ad}\pi_{bc}-\delta_{bc}\pi_{ad}+\delta_{ac}\pi_{bd}- (\delta_{ac}\delta_{bd}-\delta_{ad}\delta_{bc})\pi\right]
\ee


\begin{thebibliography}{99}



\bibitem{Deser:1976iy} 
  S.~Deser, C.~Teitelboim,
  ``Duality Transformations of Abelian and Nonabelian Gauge Fields,''
  Phys.\ Rev.\ D {\bf 13}, 1592 (1976).
  
  \bibitem{Deser}  S. Deser,
  ``Off-shell electromagnetic duality invariance,"
  J. Phys. A: Math. Gen. {\bf 15}, 1053 (1982).
  
\bibitem{Henneaux:1988gg} 
  M.~Henneaux and C.~Teitelboim,
  ``Dynamics Of Chiral (selfdual) p-Forms,''
  Phys.\ Lett.\ B {\bf 206}, 650 (1988).
  
  \bibitem{ScSe} 
  J.~H.~Schwarz and A.~Sen,
  ``Duality symmetric actions,''
  Nucl.\ Phys.\  B {\bf 411}, 35 (1994)
  [arXiv:hep-th/9304154].


\bibitem{Deser:1997mz}
  S.~Deser, A.~Gomberoff, M.~Henneaux and C.~Teitelboim,
  ``Duality, self-duality, sources and charge quantization in abelian  N-form
  theories,''
  Phys.\ Lett.\  B {\bf 400}, 80 (1997)
  [arXiv:hep-th/9702184].


  
\bibitem{Bunster:2011aw} 
  C.~Bunster and M.~Henneaux,
  ``Sp(2n,R) electric-magnetic duality as off-shell symmetry of interacting electromagnetic and scalar fields,''
  PoS HRMS {\bf 2010}, 028 (2010)
  [arXiv:1101.6064 [hep-th]].

\bibitem{Bunster:2011qp} 
  C.~Bunster and M.~Henneaux,
  ``The Action for Twisted Self-Duality,''
  Phys.\ Rev.\ D {\bf 83}, 125015 (2011)
  [arXiv:1103.3621 [hep-th]].
  
\bibitem{Henneaux:2004jw}
  M.~Henneaux and C.~Teitelboim,
  ``Duality in linearized gravity,''
  Phys.\ Rev.\  D {\bf 71}, 024018 (2005)
  [arXiv:gr-qc/0408101].
  
\bibitem{Deser:2004xt} 
  S.~Deser and D.~Seminara,
  ``Duality invariance of all free bosonic and fermionic gauge fields,''
  Phys.\ Lett.\ B {\bf 607}, 317 (2005)
  [hep-th/0411169].
  
\bibitem{Julia:2005ze} 
  B.~Julia, J.~Levie and S.~Ray,
  ``Gravitational duality near de Sitter space,''
  JHEP {\bf 0511}, 025 (2005)
  [hep-th/0507262]; \\
  B.~L.~Julia,
  ``Electric-magnetic duality beyond four dimensions and in general relativity,''
  hep-th/0512320.
  
\bibitem{Hillmann:2009zf}
  C.~Hillmann,
  ``E7(7) invariant Lagrangian of d=4 N=8 supergravity,''
  JHEP {\bf 1004}, 010 (2010)
  [arXiv:0911.5225 [hep-th]].
  
\bibitem{Bunster:2012jp} 
  C.~Bunster and M.~Henneaux,
  ``Supersymmetric electric-magnetic duality as a manifest symmetry of the action for super-Maxwell theory and linearized supergravity,''
  Phys.\ Rev.\ D {\bf 86}, 065018 (2012)
  [arXiv:1207.1761 [hep-th]].
  
  \bibitem{Comment} The literature on duality is huge. We have mentioned here only the works that recognize it as a bone fide symmetry of the action (and not just of the equations of motion) since off-shell duality  is the central feature investigated in this article.
  
 \bibitem{Others}
 Actions which contain additional fields and additional gauge symmetries and which are manifestly duality and Lorentz invariant have been proposed  in \\
  P.~Pasti, D.~P.~Sorokin and M.~Tonin,
  ``Note on manifest Lorentz and general coordinate invariance in duality
  symmetric models,''
  Phys.\ Lett.\  B {\bf 352}, 59 (1995)
  [arXiv:hep-th/9503182]; \\
  P.~Pasti, D.~P.~Sorokin and M.~Tonin,
  ``Duality symmetric actions with manifest space-time symmetries,''
  Phys.\ Rev.\  D {\bf 52}, 4277 (1995)
  [arXiv:hep-th/9506109]; \\
  I.~A.~Bandos, N.~Berkovits and D.~P.~Sorokin,
  ``Duality-symmetric eleven-dimensional supergravity and its coupling to
  M-branes,''
  Nucl.\ Phys.\  B {\bf 522}, 214 (1998)
  [arXiv:hep-th/9711055]; \\
  G.~Dall'Agata, K.~Lechner and D.~P.~Sorokin,
  ``Covariant actions for the bosonic sector of D = 10 IIB supergravity,''
  Class.\ Quant.\ Grav.\  {\bf 14}, L195 (1997)
  [arXiv:hep-th/9707044];\\
  G.~Dall'Agata, K.~Lechner and M.~Tonin,
  ``D = 10, N = IIB supergravity: Lorentz-invariant actions and duality,''
  JHEP {\bf 9807}, 017 (1998)
  [arXiv:hep-th/9806140].\\
 These actions are however non-polynomial, even when the interactions are switched off. To get a polynomial action (quadratic in the absence of interactions), one must fix the new gauge symmetry in a way that breaks Lorentz invariance.  The situation is similar to that encountered in \\
L.~Brink, M.~Henneaux and C.~Teitelboim,
  ``Covariant Hamiltonian formulation of the superparticle,''
  Nucl.\ Phys.\  B {\bf 293}, 505 (1987),\\
where  a Lorentz-invariant formulation of the Hamiltonian dynamics of the superparticle was developed but at the price of a formalism that was found to be  ``rather involved".  Again in that case, the non-manifest Lorentz invariant formulation remains by far the simplest one. 
 
  
\bibitem{Bunster:2012hm} 
  C.~Bunster and M.~Henneaux,
  ``Duality invariance implies Poincare invariance,''
  Phys.\ Rev.\ Lett.\  {\bf 110}, 011603 (2013)
  [arXiv:1208.6302 [hep-th]].


\bibitem{Julia:1982gx} 
  B.~Julia,
  ``Kac-moody Symmetry Of Gravitation And Supergravity Theories,''
  Proc. AMS-SIAM Summer Seminar on Applications of Group Theory
in Physics and Mathematical Physics, Chicago 1982, LPTENS preprint
82/22, eds. M. Flato, P. Sally and G. Zuckerman, Lectures in Applied
Mathematics, {\bf 21} (1985) 335; \\
  ``Dualities in the classical supergravity limits: Dualizations, dualities and a detour via (4k+2)-dimensions,''
  In *Cargese 1997, Strings, branes and dualities* 121-139
  [hep-th/9805083].

\bibitem{West:2001as} 
  P.~C.~West,
  ``E(11) and M theory,''
  Class.\ Quant.\ Grav.\  {\bf 18}, 4443 (2001)
  [hep-th/0104081].
  
  \bibitem{Damour:2002cu} 
  T.~Damour, M.~Henneaux and H.~Nicolai,
  ``E(10) and a 'small tension expansion' of M theory,''
  Phys.\ Rev.\ Lett.\  {\bf 89}, 221601 (2002)
  [hep-th/0207267].

\bibitem{Bunster:2012km} 
  C.~Bunster, M.~Henneaux and S.~Hortner,
  ``Gravitational Electric-Magnetic Duality, Gauge Invariance and Twisted Self-Duality,''
  J. Phys. A: Math. Theor. {\bf 46}  214016 (2013)
 [ arXiv:1207.1840 [hep-th]].
  


\bibitem{Cremmer:1998px}
  E.~Cremmer, B.~Julia, H.~Lu and C.~N.~Pope,
  ``Dualisation of dualities. II: Twisted self-duality of doubled fields  and
  superdualities,''
  Nucl.\ Phys.\  B {\bf 535}, 242 (1998)
  [arXiv:hep-th/9806106].
  

  
  \bibitem{Coho}  The cohomological theorems underlying these results go back to \\
P.J. Olver, ``Differential Hyperforms", University of Minnesota, Mathematics Report 82-101 (1982) \\
and have been analyzed in the present context of field theories with tensor fields of mixed symmetry type in \cite{DVH} and \cite{XBNB}.

\bibitem{DVH}
M.~Dubois-Violette and M.~Henneaux,
  ``Generalized cohomology for irreducible tensor fields of mixed Young symmetry type,''
  Lett.\ Math.\ Phys.\  {\bf 49}, 245 (1999)
  [math/9907135]; \\
M.~Dubois-Violette and M.~Henneaux,
  ``Tensor fields of mixed Young symmetry type and N complexes,''
  Commun.\ Math.\ Phys.\  {\bf 226}, 393 (2002)
  [math/0110088 [math-qa]].


\bibitem{XBNB}  X.~Bekaert and N.~Boulanger,
  ``Tensor gauge fields in arbitrary representations of GL(D,R): Duality and Poincare lemma,''
  Commun.\ Math.\ Phys.\  {\bf 245}, 27 (2004)
  [hep-th/0208058].

    
\bibitem{Curtright:1980yk} 
  T.~Curtright,
  ``Generalized Gauge Fields,''
  Phys.\ Lett.\ B {\bf 165}, 304 (1985).

  
\bibitem{Aulakh:1986cb} 
  C.~S.~Aulakh, I.~G.~Koh and S.~Ouvry,
  ``Higher Spin Fields With Mixed Symmetry,''
  Phys.\ Lett.\ B {\bf 173}, 284 (1986).
  
\bibitem{Labastida:1986gy} 
  J.~M.~F.~Labastida and T.~R.~Morris,
  ``Massless Mixed Symmetry Bosonic Free Fields,''
  Phys.\ Lett.\ B {\bf 180}, 101 (1986).
  
 \bibitem{Hull:2000zn} 
  C.~M.~Hull,
  ``Strongly coupled gravity and duality,''
  Nucl.\ Phys.\ B {\bf 583}, 237 (2000)
  [hep-th/0004195].
  
   
\bibitem{Boulanger:2003vs} 
  N.~Boulanger, S.~Cnockaert and M.~Henneaux,
  ``A note on spin s duality,''
  JHEP {\bf 0306}, 060 (2003)
  [hep-th/0306023].
  
  
\bibitem{Bakas:2009da} 
  I.~Bakas,
  ``Dual photons and gravitons,''
  arXiv:0910.1739 [hep-th].
  

\bibitem{Hull:2001iu} 
  C.~M.~Hull,
  ``Duality in gravity and higher spin gauge fields,''
  JHEP {\bf 0109}, 027 (2001)
  [hep-th/0107149].
  
  

\bibitem{Dirac} P.~A.~M.~Dirac,
  ``The Theory of gravitation in Hamiltonian form,''
  Proc.\ Roy.\ Soc.\ Lond.\ A {\bf 246}, 333 (1958); \\
  P.~A.~M.~Dirac,
  ``Fixation of coordinates in the Hamiltonian theory of gravitation,''
  Phys.\ Rev.\  {\bf 114}, 924 (1959).


\bibitem{Boulanger:2008nd} 
  N.~Boulanger and O.~Hohm,
  ``Non-linear parent action and dual gravity,''
  Phys.\ Rev.\ D {\bf 78}, 064027 (2008)
  [arXiv:0806.2775 [hep-th]].
  
     
  
\bibitem{Senovilla:1999xz}
J.~M.~M.~Senovilla,
``Superenergy tensors,''
Class.\ Quant.\ Grav.\ {\bf 17}, 2799 (2000)
[gr-qc/9906087];\\
J.~M.~M.~Senovilla,
``General electric magnetic decomposition of fields, positivity and Rainich-like conditions,''
gr-qc/0010095; \\
S.~Hervik, M.~Ortaggio and L.~Wylleman,
  ``Minimal tensors and purely electric or magnetic spacetimes of arbitrary dimension,''
  arXiv:1203.3563 [gr-qc]; \\
    S.~Hervik, M.~Ortaggio and L.~Wylleman,
  ``Electric and magnetic Weyl tensors in higher dimensions,''
  arXiv:1301.3691 [gr-qc].

\bibitem{Bunster:2013tc} 
  C.~Bunster, M.~Henneaux and S.~Hortner,
  ``Duality-invariant bimetric formulation of linearized gravity,''
  arXiv:1301.5496 [hep-th].
  
\bibitem{Bergshoeff:2012yz} 
  E.~A.~Bergshoeff, M.~Kovacevic, J.~Rosseel and Y.~Yin,
  ``On Topologically Massive Spin-2 Gauge Theories beyond Three Dimensions,''
  JHEP {\bf 1210}, 055 (2012)
  [arXiv:1207.0192 [hep-th]].
 
\bibitem{Kallosh:2011dp} 
  R.~Kallosh,
  ``$E_{7(7)}$ Symmetry and Finiteness of N=8 Supergravity,''
  JHEP {\bf 1203}, 083 (2012)
  [arXiv:1103.4115 [hep-th]]; \\
    R.~Kallosh,
  ``$N=8$ Counterterms and $E_{7(7)}$ Current Conservation,''
  JHEP {\bf 1106}, 073 (2011)
  [arXiv:1104.5480 [hep-th]].
 
\bibitem{Gunaydin:2013pma} 
  M.~Gunaydin and R.~Kallosh,
  ``Obstruction to $E_{7(7)} $ Deformation in N=8 Supergravity,''
  arXiv:1303.3540 [hep-th].








  
  
\end{thebibliography}
\end{document}